\documentclass{article}
\usepackage[letterpaper, margin=1in]{geometry}
\usepackage{amsmath}
\usepackage{graphicx}
\usepackage{authblk}
\usepackage{url}
\usepackage{xcolor}
\usepackage{xfrac}
\usepackage{pdflscape}
\usepackage{hyperref}

\newcommand{\ellamp}{\boldsymbol{\varphi}}
\newcommand{\ellpar}{\boldsymbol{m}}

\definecolor{mybcolor}{RGB}{247, 87, 69}
\definecolor{myalphacolor}{RGB}{82, 170, 41} % a bit darker
\definecolor{mybetacolor}{RGB}{153, 183, 29} % a bit darker
\definecolor{mythetacolor}{RGB}{34, 168, 132}
\definecolor{myDeltacolor}{RGB}{42, 120, 142}
\definecolor{mynucolor}{RGB}{65, 68, 135}
\definecolor{mynormalcolor}{RGB}{153, 102, 51}
\definecolor{mypsicolor}{RGB}{255, 102, 0}

\definecolor{mypcolor}{RGB}{255,85,153}
\definecolor{myucolor}{RGB}{221, 85, 255}
\definecolor{myvcolor}{RGB}{204, 0, 255}
\definecolor{mywcolor}{RGB}{113, 55, 200}
\definecolor{myzcolor}{RGB}{137, 44, 160}

\definecolor{mylinkcolor}{RGB}{5, 70, 5}
\definecolor{mycitecolor}{RGB}{100, 5, 5}
\definecolor{myurlcolor}{RGB}{5, 5, 100}

\hypersetup{
    colorlinks=true,
    linkcolor=mylinkcolor,
    citecolor=mycitecolor,
    filecolor=magenta,
    urlcolor=myurlcolor,
}

\newcommand{\bc}{\textcolor{mybcolor}{\hat{B}}}
\newcommand{\bfield}{\vec{B}}
\newcommand{\alphac}{\textcolor{myalphacolor}{\boldsymbol{\alpha}}}
\newcommand{\betac}{\textcolor{mybetacolor}{\boldsymbol{\beta}}}
\newcommand{\thetac}{\textcolor{mythetacolor}{\theta}}
\newcommand{\Deltac}{\textcolor{myDeltacolor}{\Delta}}
\newcommand{\nuc}{\textcolor{mynucolor}{\nu}}
\newcommand{\normc}{\textcolor{mynormalcolor}{\hat{n}}}
\newcommand{\psic}{\textcolor{mypsicolor}{\boldsymbol{\psi}}}

\newcommand{\pc}{\textcolor{mypcolor}{p}}
\newcommand{\uc}{\textcolor{myucolor}{u}}
\newcommand{\vc}{\textcolor{myvcolor}{v}}
\newcommand{\wc}{\textcolor{mywcolor}{w}}
\newcommand{\zc}{\textcolor{myzcolor}{z}}

\newcommand{\anar}{\textit{anarr\'{i}ma}}
\newcommand{\Anar}{\textit{Anarr\'{i}ma}}

\usepackage{tabularx} % For better control over table column widths

\begin{document}

\title{Analytic neutron wall loading from spin-polarized fusion in axisymmetric geometries}

\date{\today}
\author[1,2]{Jacob A. Schwartz}
\affil[1]{Previous affiliation: Princeton Plasma Physics Laboratory, Princeton NJ, USA}
\affil[2]{Present affiliation: Marathon Fusion, San Francisco, CA, USA}

\maketitle

\vspace{1em}

% \begin{flushleft}
%   J.A. SCHWARTZ \\
%   Princeton Plasma Physics Laboratory \\
%   Princeton, NJ, United States of America \\
%   Email: jschwartz@pppl.gov
% \end{flushleft}

\section*{Abstract}
{\fontsize{9pt}{12pt}\selectfont\hspace{1cm} 

Spin-polarized nuclei can be used to enhance the fusion reaction rate, preferentially direct neutrons and alpha particles, or do some combination of these.
In this paper we present formulas for the neutron wall load (NWL) on the walls of an axisymmetric torus with a convex cross section from filamentary ring sources with arbitrary fuel polarizations and arbitrary local magnetic field directions.
While the NWL does not include neutron scattering, which would require a more sophisticated treatment, these formulas can be evaluated very quickly.
This can help a fusion machine designer build intuition about the effects of spin-polarization on neutron wall loading.
The formulas are also fully differentiable, so they could be included in a fusion systems optimization code. 
As an example, we present sets of first wall shapes optimized to have a uniform NWL for each of the main polarization modes.
}

\renewcommand{\footnoterule}{\hrule width \linewidth}

\section{Introduction}

Polarizing the nuclei of fusion fuel prior to their injection into a plasma could potentially allow for smaller, higher-performing, and less-expensive fusion devices\cite{parisiSimultaneousEnhancementTritium2024}.
Two phenomena associated with spin-polarized fusion reactions might be leveraged: first, an enhancement in the fusion rate by up to 50\%, and second, preferred directions for the emission of neutrons and alpha particles.
In particular, the latter could be used to direct neutrons away from more sensitive reactor structures---the center stack and divertors in a tokamak, the end plugs in a mirror machine, or the flyer ring in a levitated dipole---and toward more robust outer walls.
At the same time, alpha particles (directed oppositely to the neutrons by conservation of momentum) could be given preferentially parallel or perpendicular momenta. This could be used to enhance core heating or reduce end losses.
The polarizations that produce an enhanced reaction rate are not independent from the ones which affect directionality.
Therefore, the optimal choice will need to be made in a system context.

While more detailed reactor studies use sophisticated Monte Carlo neutron transport codes\cite{romanoOpenMCStateoftheartMonte2015, kuleszaMCNPCodeVersion2022} to compute volumetric heating in the first wall and blanket\cite{el-guebalyARIESSTNuclearAnalysis2003,raffrayEngineeringDesignAnalysis2008, fausserTokamakDTNeutron2012}, scoping studies often use simplified `neutron wall load' (NWL) calculations to estimate the flux from the plasma directly impinging on different parts of the first wall\cite{lionDeterministicMethodFast2022}.
This paper presents formulas for the NWL from polarized sources in various axisymmetric reactor geometries.
In each case, we consider infinitesimal ring plasma sources.
For some of the geometries presented, these can have arbitrary local $\vec{B}$-field directions.
The formulas can be summed to allow for arbitrary polarization mixes.
While they can be lengthy, they are in terms of analytic expressions which are quick to compute using standard special function libraries and are automatically differentiable.
The latter would allow them to be part of an efficient gradient-based optimization scheme.
For example in a tokamak geometry one could modify the plasma shape and $q$ profile, first wall shape, and polarization mix in order to optimize an objective such as the total fusion power, while constraining quantities like the maximum wall load or the ratio of maximum to average wall load, often called the ``peaking factor''\cite{woodsNeutronWallLoading1977, davisNeutronicsAspectsFESSFNSF2018}.

Some important limitations apply to this work.
First, it does not treat systems where the center-of-mass velocity of the reactants is comparable to the neutron velocity, such as MeV-energy beam-driven processes.
Second, with one exception we only treat reactors with convex poloidal cross sections, such as a simplified tokamak geometry.
Practical devices often have non-convex cross sections, particularly around the divertor.

\subsection{Comparison to previous work}
Neutron wall load calculations which do not use a Monte Carlo method are often performed via ray tracing through the plasma.
For example, Chapin and Price\cite{chapinComparisonDeuteriumTritiumNeutron1976} trace rays from a point on the wall through the plasma until it hits another part of the torus wall.
Chan et.\ al.\cite{chanPhysicsBasisFusion2010} and Yang et al.~\cite{yangEffectsNeutronSource2024} use a `multifilament' approach similar to the one described here, but it is not clear how the limits of integration for each filament are determined, and the toroidal integral over each filament is performed numerically.
This work focuses on an analytic evaluation of the toroidal integral.
Lion\cite{lionDeterministicMethodFast2022} computes the neutron wall load in a stellarator geometry, which necessitates numerical integration due to the 3D nature of the source.

\subsection{Sections in this paper}
Section~\ref{ssec:physics} briefly describes the relevant physics of spin-polarized DT fusion and section~\ref{ssec:nwldef} defines the `neutron wall load' mathematically.
Section~\ref{sec:rectexample} presents results for a simplified tokamak system, in order to give the reader intuition about the directional effects of neutrons from polarized fuel.
Section~\ref{sec:rectanglemak} presents geometries, integrals, and evaluates the integrals for a rectangular tokamak geometry.
Section~\ref{sec:python} briefly describes \anar{}, a new open-source python package which implements the formulas in this work and is used for computing the examples.
Section~\ref{sec:arbtorus} presents the extensions needed for a more general convex-cross section tokamak geometry.
In a brief diversion to a different fusion machine geometry, Section~\ref{sec:dipole} provides formulas required for the exterior part a levitated dipole's flyer ring.
Finally, we summarize and suggest future work.

\subsection{Physics of the spin-polarized DT fusion reaction}
\label{ssec:physics}

The papers by Kulsrud et al.\ \cite{kulsrudFusionReactorPlasmas1982} as well as the book edited by Ciullo et al.\ \cite{ciulloNuclearFusionPolarized2016} are good references for the physics of spin-polarized fusion; this section begins following material from the first page of the former.
The deuterium nucleus has total spin 1, so relative to the magnetic field direction it can have three states: $+1, 0$, and $-1$. These are referred to as parallel, transverse, and antiparallel spin projections.
The tritium nucleus has total spin $\sfrac{1}{2}$ so its spin projection $m_z$ can be $\sfrac{1}{2}$ or $-\sfrac{1}{2}$: parallel or antiparallel.

Let the fractions of colliding deuterium ions in each state be denoted $d_+, d_0, d_-$ and for tritium ions $t_+, t_-$.
When two nuclei meet there are therefore six possible combinations of their spin projections.
Symmetry around the magnetic field direction reduces this to three distinct combinations, and the fraction of nuclei which collide in each mode are termed $a$, $b$ and $c$:

\begin{equation}
\begin{aligned}
a &= d_+ t_+ + d_{-} t_{-} \\
b &= d_0 = d_0 t_+ + d_0 t_- \\
c &= d_+ t_- + d_{-} t_{+}.
\end{aligned}
\end{equation}
By construction these sum to unity: the mix of spins has three independent choices, but the resulting combination of modes has only two degrees of freedom.
For a non-polarized plasma, $a=b=c=1/3$.

The differential cross section for neutron emission (and alpha particle emission) is
\begin{equation}
\frac{d\sigma}{d\Omega} = \frac{\sigma_0}{2\pi}\left(\frac{3}{4}a \sin^2\theta + \left(\frac{2}{3}b + \frac{1}{3} c \right)\left(\frac{1}{4} + \frac{3}{4} \cos^2\theta\right)\right)
\label{eq:kulsrud}
\end{equation}
where $\sigma_0$ is the differential cross section of fusion in the $S=\sfrac{3}{2}$ combined spin state and
$\theta$ is the angle relative to the local magnetic field direction $\hat{B}$.
For non-polarized fusion, the integrated cross section is $2\sigma_0/3$ and the differential cross section is independent of $\theta$.
If $a=1$ (`A mode') the integrated cross section is $\sigma_0$---enhanced by 50\% relative to the non-polarized state---and neutrons emerge with a distribution proportional to $\sin^2\theta$.
This mode could be used to enhance the cross section and produce neutrons and alphas preferentially perpendicular to the field.
If $b=1$ (`B mode'), the integrated cross section is unchanged from the non-polarized case, at $2\sigma_0/3$, and neutrons are directed preferentially parallel to the field, with an angular distribution proportional to $\sfrac{1}{4} + \sfrac{3}{4}\cos^2\theta$.
This mode reduces the neutron and alpha intensity perpendicular to the field, and enhances it parallel to the field.
If $c=1$ (`C mode') the directionality is the same as for the B mode, but the integrated cross section is \textit{reduced} by 50\% to $\sigma_0/3$.

\begin{figure}[ht]
    \centering
    \includegraphics[width=0.60\textwidth]{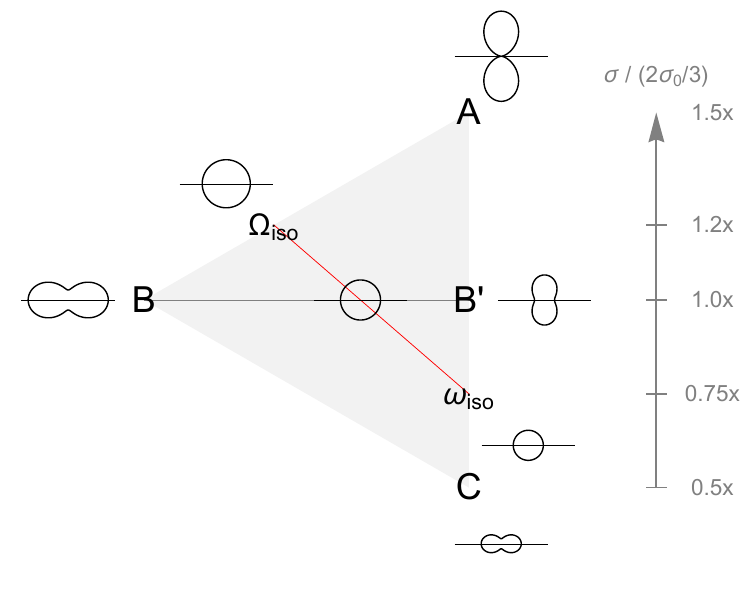}
	\caption{Differential cross sections and total cross sections for possible combinations of polarization modes $(a, b, c)$.
	The corners are pure A, B, and C, and any point in the triangle is a possible combination.
	The total cross section increases upward, indicated by the vertical axis: values are relative to the non-polarized case.
	The red line shows combinations producing isotropic emission.
	Small polar plots for each labeled mode show the differential cross section as a function of the angle from the (horizontal) magnetic field direction.
	Note that the relative size of these plots can be misleading: the angular emission strength is proportional to the \textit{linear} distance from each plots' origin, not their areas, nor volumes of the solids implied by rotation around the magnetic field direction.
	}
    \label{fig:spintriangle}
\end{figure}

Figure~\ref{fig:spintriangle} illustrates the differential cross sections and total cross sections of the polarization modes and their linear combinations.
Aside from the non-polarized case and pure-A, B, and C modes mentioned above, three other modes are highlighted. 
First, the $\Omega_\mathrm{iso}$ `maximal isotropic' mode, $\sfrac{2}{5}\,\mathrm{A} + \sfrac{3}{5}\,\mathrm{B}$, has a total cross section of 1.2 of the non-polarized case.
This could enhance the reaction rate without generating anisotropic $\alpha$ particle distributions (which could couple to certain plasma instabilities), or for maximizing the reaction rate with a uniform neutron wall load in a spherical ICF chamber.
Second, the $\omega_\mathrm{iso}$ `minimal isotropic' mode, $\sfrac{1}{4}\,\mathrm{A} + \sfrac{3}{4}\,\mathrm{C}$, has a total cross section of 0.75 of the non-polarized case.
The red line from $\Omega_\mathrm{iso}$ to $\omega_\mathrm{iso}$ is the set of combinations with isotropic emission.
Finally, the B' or `anti-B' mode is noted here because it is mentioned in other literature: Heidbrink \cite{heidbrinkResearchProgramMeasure2024} refers to it as the Tensor D mode.

\subsection{Definition of neutron wall load}
\label{ssec:nwldef}
Following Lion\cite{lionDeterministicMethodFast2022}, the neutron wall load from an isotropic neutron source is the energy per neutron multiplied by the neutron current through a patch of the first wall:
\begin{equation}
	Q^{\mathrm{NWL}}(\boldsymbol{r}_w) = \frac{E_n}{4\pi}\int_{V_s}\,dV\,\left(\Phi^\mathrm{source}\;\hat{n}\cdot\frac{\boldsymbol{r}_s - \boldsymbol{r}_w}{|\boldsymbol{r}_s - \boldsymbol{r}_w|^3}\right)
\end{equation}
where $E_n$ is the energy per neutron, $V_s$ the volume of the plasma, $\Phi^{\mathrm{source}}(\boldsymbol{r}_s)$ is the local neutron source density at a plasma source point $\boldsymbol{r}_s$, and $\hat{n}(\boldsymbol{r}_w)$ is the normal to the first wall at a wall point $\boldsymbol{r}_w$.
The factor $1/(4\pi)$ is the solid angle over which particles are emitted.
This factor is constant for an isotropic source;
as described in the subsection above, a spin-polarized emission distribution has an additional term in the integrand which depends on the angle $\theta$ between the local magnetic field direction $\hat{B}$ and $(\boldsymbol{r}_s - \boldsymbol{r}_w)$.

This work emphasizes the geometric aspects of the calculation rather than a high-fidelity physical model, so we assume that $E_n$ is the same for every neutron\footnote{This ignores the energy spread from the ion thermal velocity, beam-target fusion, or knock-on fusion effects\cite{kallneObservationAlphaParticle2000}};
we therefore drop $E_n$ from formulas.
As a second simplication, rather than consider volumetric neutron sources $\Phi^{\mathrm{source}}(\boldsymbol{r}_s)$ with possible 3D variation, we consider filamentary ring sources with emission rates of 1 per unit length per unit time.
These can be combined with a weighted sum to approximate an more complex axisymmetric source.

%Note that some authors (such as Chapin~\cite{chapinComparisonDeuteriumTritiumNeutron1976}) use `neutron wall load' to refer also to the angular distribution
%
Formulas for the neutron wall load from a filamentary ring will be derived and presented in Section~\ref{sec:rectanglemak}.
First, Section~\ref{sec:rectexample} provides a preview of the effect of spin polarization on the (normalized) wall loads in a tokamak-like geometry.

% Following Chapin and Price\cite{chapinComparisonDeuteriumTritiumNeutron1976}, the neutron wall load is the fusion neutron current through a patch of the first wall multiplied by the energy per neutron.
% The current through a patch is defined as 
% \begin{equation}
% 	J(\boldsymbol{r}_w) = \int_{-1}^{+1} \mu F(\boldsymbol{r}_w, \mu) \,d\mu,
% \label{eq:neutroncurrent}
% \end{equation}
% where $\boldsymbol{r}_w$ is a point on the wall, $\mu = \cos\nu$ is the cosine of the angle between the neutron's direction and the normal to the patch, and $F(\boldsymbol{r}_w, \mu)$ is the `polar flux' distribution at $\boldsymbol{r}_w$.
% In this model we consider only neutrons travelling along lines of sight directly from the plasma so $F(\boldsymbol{r}_w, \mu <=0) = 0$.

\section{Example: NWL from polarized fuels in a square-cross-section torus}
\label{sec:rectexample}

\begin{figure}[htb]
    \centering
    \includegraphics[width=0.65\textwidth]{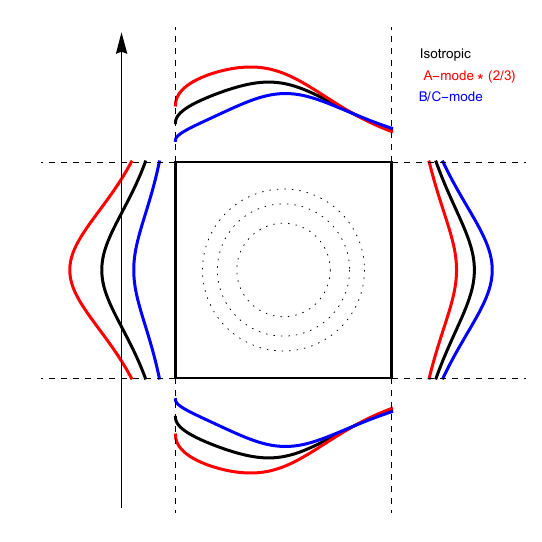}
    \caption{Relative neutron wall load intensity patterns on the four walls of an axisymmetric square-cross-section torus from an example circular plasma in a toroidal field. The black, red, and blue traces show the local intensity from isotropic emission, the A mode and the B/C modes, respectively. The walls (thick square) are baselines for each plot. The intensity from the A mode has been multiplied by $2/3$ to emphasize the directionality of neutrons under a constant fusion rate.
     Notably, the B/C modes reduces the neutron wall load on the center stack by 43\%.}
    \label{fig:rectanglemak}
\end{figure}

Polarization can significantly alter the neutron wall load.
This section demonstrates the neutron wall load (NWL) from different polarization modes in a square-cross-section tokamak-like geometry.
For simplicity we use a circular plasma with aspect ratio $2.5$ in a square torus, shown in Figure~1.
The local fusion reaction density is parabolic in the normalized minor radius $\rho$, proportional to $(1 - \rho^2)$.
We show the NWL for the isotropic non-polarized case, for $a=1$ and for $b + c = 1$ (`B/C modes').
In order to emphasize changes in directionality rather than the changing cross section, the total fusion rate is held constant in the three cases.

Even with non-polarized fusion (isotropic neutron emission) the outer wall has a higher intensity than the inner wall by about 12\%; this is due to purely geometric effects.
The A mode increases the intensity on the inboard side (by 43\% at the midplane), as well as on the top and bottom, and decreases (by 22\% at the midplane) the intensity on the outboard side.
The B/C modes have the opposite effects.
They reduce the intensity on the inboard (by the same 43\% at the midplane), top and bottom, and increases it (by the same 22\% at midplane) on the outboard side.
It is somewhat counterintuitive that the intensity changes are much larger on the inner wall than the outboard wall; this is possible due to the smaller area on the inboard wall.

The lower NWL on the center stack in B/C modes could be particularly useful for spherical tokamaks, where the center stack is a critical component.
The lower NWL there would straightforwardly decrease heating loads and increase magnet lifetimes.
The lower loads could potentially be leveraged to use thinner shields or eschew the blanket there entirely to make the center stack even thinner and access lower aspect ratio configurations.

The next section will describe the construction of the formulas used to compute these intensities, for rectangular vessels.

\section{Interior of torus with rectangular cross section}
\label{sec:rectanglemak}

This section will show how to compute the neutron wall load intensity on the interior walls of a torus with a rectangular cross section.
These formulas provide the foundation for computing the intensity in more complex geometries with sloped walls.
We will start from a relatively simple case and add complications one at a time.
First we will present the intensity on the inboard wall (or center stack) from isotropic emission.
Next will be the intensity from different polarization patterns, but assuming a purely toroidal field.
Other than isotropic emission, the important polarization patterns are
\begin{itemize}
  \item $\sin^2\theta$, for the A mode,
  \item $\cos^2 \theta$, as a term of the B and C modes
  \item $1/4 + 3/4 \cos^2\theta$, for the B and C modes.
\end{itemize}
While the $\cos^2\theta$ mode is not important from a physics perspective, it can be used to cross-check with the $\sin^2\theta$ and isotropic modes since $\cos^2\theta + \sin^2\theta = 1$.
Finally we will compute the intensity from a ring of plasma where the local $\vec{B}$ field points in a (toroidally symmetric) arbitrary direction.
The computations for the floor/ceiling and the outboard walls are very similar to the ones for the inboard wall.
We will highlight the key differences rather than present every detail.

\subsection*{A note about quantities}

In order to make the computations a bit simpler, we will compute geometric quantities $g$ with units of $(\textrm{length})^{-1}$ which can then be related to the physical intensities $i$ with units $(\textrm{particles})\cdot(\textrm{length})^{-2}\cdot(\textrm{time})^{-1}$ by multiplying by a differential ring source with units of $(\textrm{particles})\cdot(\textrm{length})^{-1}\cdot(\textrm{time})^{-1}$.
An isotropically-emitting source with total strength $1$ per length per time will have an angular intensity of $1/4\pi$ per steradian. We ignore this $4\pi$ factor in our calculations of $g$.
We also ignore the polarization-specific intensity factors in Equation~\ref{eq:kulsrud} other than the angular factors $\sin^2\theta$ and $\sfrac{1}{4} + \sfrac{3}{4} \cos^2\theta$ mentioned above. (Note that while Eq.~\ref{eq:kulsrud} describes the differential cross section for specific collisions, the differential \textit{rate coefficient} $\left<\sigma v\right>$ will depend on the angle $\theta$ and the polarization factors $a,b,c$ in a similar way.)

\subsection*{A note about naming of the formulas and terms}

The geometric quantities $g$ will be named using subscripts.
First is one of five geometries: $Ri$, $Rf$, or $Ro$ for a rectangular torus inboard, floor/ceiling, or outboard, respectively;
for a generic horizontal and vertical components, $H$ or $V$, respectively.
Next a letter for the polarization mode: $I$ for isotropic; $A$ for A mode or $\sin^2\theta$; lowercase $c$ for $\cos^2\theta$; or $B$ for the B/C modes, $\sfrac{1}{4} + \sfrac{3}{4} \cos^2\theta$.
An optional lowercase $a$ suffix specifies the $\bfield$ field is angled not in the toroidal direction.

The $g$ expressions generally have three terms. When necessary these are referred to individually as $h$.
The naming scheme for the $h$ terms is the same, with an additional letter appended: $L$ for the leading term (without any elliptic integrals), or an $F$ or $E$ depending on which type of elliptic integral.

\subsection{Inboard side, purely toroidal field}\label{ssec:inboardToroidalField}
\begin{figure}[htb]
    \centering
    \includegraphics[width=0.75\textwidth]{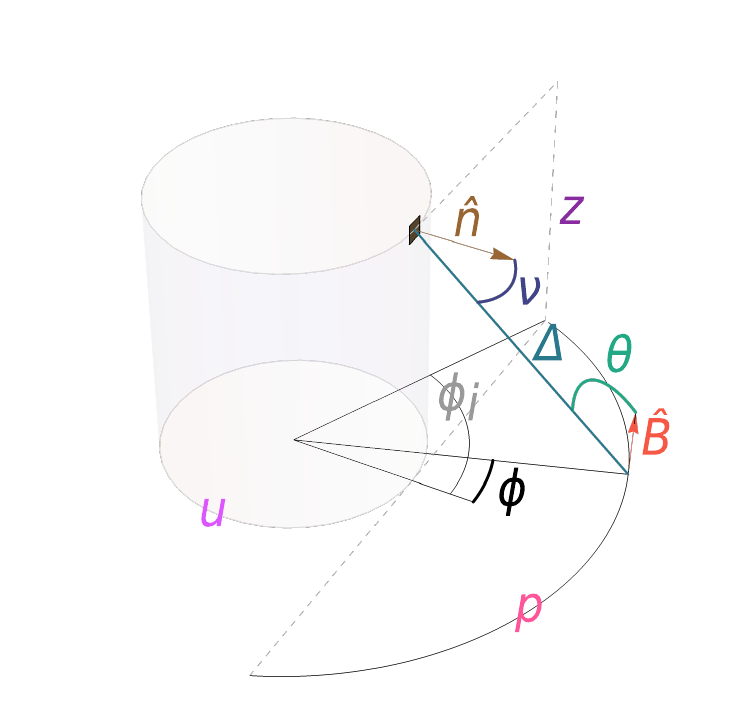}
    \caption{Geometry to compute the intensity on a target on the inboard wall at radius $\uc$, from an infinitesimal ring source of plasma at radius $\pc$ and relative height $\zc$.
    The magnetic field direction $\bc$ is purely toroidal.
    The vector between the source (at toroidal angle $\phi$) and target is $\Deltac$.
    The angle between $\bc$ and $\Deltac$ is $\thetac$, and between between $\Deltac$ and the target normal $\normc$ is $\nuc$. The plasma ring source is integrated over the range visible from the target: $\pm \phi_i$.
    }
    \label{fig:rectInboard}
\end{figure}

The geometry for the first four $g$ integrals is shown in Fig.~\ref{fig:rectInboard}.
The target is on the inner wall (or center stack), with location in Cartesian coordinates $T = (x=\uc, y=0, z=0)$.
It normal $\normc = (1, 0, 0)$ faces radially outward.
The source is a plasma ring at radius $\pc$ and height $\zc$ relative to the target.
We will integrate over the ring angle $\phi$. The limits of integration, $\pm \phi_i = \cos^{-1}(\uc/\pc)$ are the furthest extents of the ring which are visible from the target.
Each element of the ring source has location
$$ S = (\pc \cos \phi, \pc \sin \phi, \zc).$$
The vector from the source to the target is
$$\Deltac = (\pc \cos \phi - \uc, \pc \sin \phi, \zc);$$
it has length $|\Deltac| = (\pc^2 + \zc^2 + \uc^2 - 2 \pc \uc \cos \phi)^{1/2}$.
The angle between $\Deltac$ and the target's normal is $\nuc$.
The intensity on the target is proportional to the cosine of this angle,
$$\cos\nuc = \frac{\Deltac \cdot \normc}{|\Deltac|} = \frac{\pc \cos \phi - \uc}{|\Deltac|}.$$
The angle between the $\bfield$ field direction $\bc$ and $\Deltac$ is $\thetac$, which is important for polarized sources, but not for isotropic emission.

\subsubsection{Isotropic}

The reduced intensity $g$ on an element of the inboard wall from an isotropic source is

\begin{equation}
  g_{RiI} = \int_{-\phi_i}^{\phi_i} \;d\phi\; \frac{\pc \cos \nuc}{|\Deltac|^2}
  =
  \int_{-\phi_i}^{\phi_i}  \;d\phi\; \frac{p (p \cos (\phi )-u)}{\left(p^2 + u^2 + z^2 -2 p u \cos (\phi )\right)^{3/2}}.
  \label{eq:gRiIsetup}
\end{equation}
where the first $\pc$ is the Jacobian of integration in cylindrical coordinates, $\cos\nuc$ captures the angle between the source and the target's normal, and the $|\Deltac|^2$ in the denominator is the diminishment of intensity with distance.
This evalutes to
\begin{equation}
\begin{split}
  g_{RiI} &= \frac{4 p \sqrt{\left(p^2-u^2\right) \left(p^2-u^2+z^2\right)}}{q} \\
   &-\frac{2 p}{u \sqrt{(p-u)^2+z^2}}F(\ellamp, \ellpar)
    +\frac{2 p \left(p^2-u^2+z^2\right) \sqrt{(p-u)^2+z^2}
   }{q u}E(\ellamp, \ellpar)
\end{split}
\label{eq:gRiI}
\end{equation}
where $q\equiv p^4 - 2 p^2 (u^2-z^2) + (u^2 + z^2)^2$ and incomplete elliptic integrals $F(\ellamp, \ellpar)$ and $E(\ellamp, \ellpar)$ have amplitude $\ellamp = \phi_i/2$ and parameter $\ellpar = -4 p u/((p - u)^2 + z^2))$.
 
For arbitrary limits on the integral $\pm \phi_m$ instead of $\pm\phi_i$, and for arbitrary target radius $r$ in place of $u$, the leading term is
\begin{equation}
    h_{HIL} = \frac{4 p^2  \left(p^2-r^2+z^2\right)\sin \phi_m }{q_r \sqrt{p^2+r^2+z^2-2 p r \cos \phi_m }}
    \label{eq:hHIL}
\end{equation}
where $q_r$ is in terms of $r$ instead of $u$.
The corresponding terms $h_{HIF}$ and $h_{HIE}$ have the same form as those shown in Eq.~\ref{eq:gRiI}, $h_{RiIF}$ and $h_{RiIE}$; the only difference is the amplitude of the elliptic integral $\ellamp=\phi_m/2$, and of course $r$ in place of $u$. This is true of the following integrals as well; thus we only need to provide the leading `\textit{L}' terms for the more general case.
For the rest of the $g_{\ldots}$ expressions in this section, the corresponding leading $h_{\ldots L}$ terms will be given in Section~\ref{sec:arbtorus} where they are used to calculate the neutron wall load for elements with normals at any angle.

\subsubsection{Proportional to \texorpdfstring{$\sin^2\thetac$}{Sine theta squared}, A mode}

With integrals for polarized radiation patterns the $\bfield$ field direction
$\bc = (-\sin\phi, \cos\phi, 0)$
and the angle $\thetac$ between $\bc$ and $\Deltac$ becomes important.
The A mode radiation pattern is proportional to $\sin^2\thetac$.
We use the dot product to first compute $\cos\thetac$,
\begin{equation}
  \cos\thetac = \frac{\bc \cdot \Deltac}{|\Deltac|} = \frac{\uc \sin\phi}{|\Deltac|},
\end{equation}
and the quantity $\sin^2\thetac$ is computed via $1 - \cos^2\thetac$.
The reduced intensity is
\begin{equation}
  g_{RiA} = \int_{-\phi_m}^{\phi_m} \;d\phi\; \frac{\pc \cos \nuc \sin^2\thetac}{|\Deltac|^2}
  =
  \int_{-\phi_m}^{\phi_m}  \;d\phi\; \frac{p (p \cos (\phi )-u) \left(p^2 + z^2 -2 p u \cos (\phi )+u^2 \cos ^2(\phi )\right)}{\left(p^2 + u^2 + z^2 -2 p u \cos (\phi )\right)^{5/2}}.
  \label{eq:gRiAsetup}
\end{equation}
This evaluates to
\begin{equation}
  \begin{split}
      g_{RiA}&=\frac{4 \sqrt{p^2-u^2} \left(2 p^4+p^2 \left(z^2-2 u^2\right)-z^2 \left(u^2+z^2\right)\right)}{3 p q \sqrt{p^2-u^2+z^2}} \\
    &+\frac{2 \left(-p^2+u^2+2 z^2\right)}{3 p u \sqrt{(u-p)^2+z^2}}F
     -\frac{2 \sqrt{(p-u)^2+z^2} \left(-p^4+u^4+\left(p^2+3 u^2\right) z^2+2 z^4\right) }{3 p q u}E
  \end{split}
  \label{eq:gRiA}
\end{equation}
where here and for the rest of this subsection we elide the argument and parameter of the elliptic integral; they are the same as for the isotropic case.
\subsubsection{Proportional to \texorpdfstring{$\cos^2\thetac$}{Cosine theta squared}}
The reduced intensity is
\begin{equation}
  g_{Ric} = \int_{-\phi_i}^{\phi_i} \;d\phi\; \frac{\pc \cos \nuc \cos^2\thetac}{|\Deltac|^2}
  =
  \int_{-\phi_i}^{\phi_i}  \;d\phi\; \frac{p u^2 \sin ^2(\phi ) (p \cos (\phi )-u)}{\left((\cos (\theta )-\rho )^2+\sin ^2(\theta )+z^2\right)^{5/2}}.
  \label{eq:gRicsetup}
\end{equation}
This evaluates to
\begin{equation}
  \begin{split}
    g_{Ric} &= \frac{4 \sqrt{p^2-u^2} \left(q-u^2 \left(-p^2+u^2+z^2\right)\right)}{3 p q \sqrt{p^2-u^2+z^2}} \\
    &-\frac{2 \left(2 p^2+u^2+2 z^2\right) }{3 p u \sqrt{(p-u)^2+z^2}}F\\
 &+\frac{2 \left(2 q-u^2 \left(-p^2+u^2+z^2\right)\right)}{3 p u \sqrt{(p-u)^2+z^2} \left((p+u)^2+z^2\right)} E.
  \end{split}
  \label{eq:gRic}
\end{equation}
Since they are not applicable to any physical emission pattern, after this point in the paper expressions for reduced intensities proportional to $\cos^2\thetac$ are listed in Appendix~\ref{sec:cosdependence}.
\subsubsection{B/C modes}
We combine the corresponding closed-form, $F$, and $E$ terms from the isotropic and $\cos^2\thetac$ expressions:
$$g_{RiB} = \frac{1}{4}g_{RiI} + \frac{3}{4}g_{Ric}.$$
This yields
\begin{equation}
  \begin{split}
    g_{RiB} &= \frac{\sqrt{p^2-u^2} \left(2 p^2 (p^2-u^2)+\left(3 p^2+u^2\right) z^2+z^4\right)}{p q \sqrt{p^2-u^2+z^2}} \\
    &-\frac{3 p^2+u^2+2 z^2}{2 p u \sqrt{(p-u)^2+z^2}}F
   +\frac{\left(3 p^4+u^4+3 u^2 z^2+2 z^4+p^2 \left(-4 u^2+5 z^2\right)\right)}{2 p u \sqrt{(p-u)^2+z^2} \left((p+u)^2+z^2\right)}E.
  \end{split}
  \label{eq:gRiB}
\end{equation}

\subsection{Angled field}\label{ssec:inboardAngledField}
\begin{figure}[htb]
    \centering
    \includegraphics[width=0.50\textwidth]{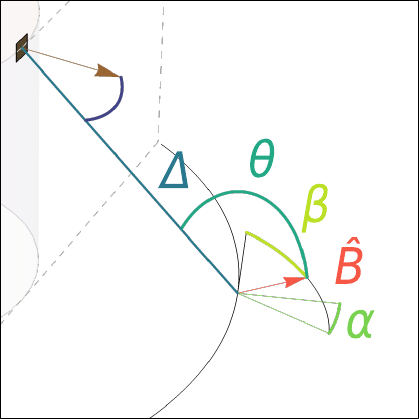}
	\caption{Definitions for the angled $\vec{B}$-field complication.
    This adds a detail to the construction of $\thetac$ in the overall geometry of Figure~\ref{fig:rectInboard}.
    The angle $\betac$ is between the magnetic field direction $\bc$ and the positive toroidal direction.
    The angle $\alphac$ specifies the plane that $\betac$ lies in.
    }
    \label{fig:rectInboardArbitraryAngle}
\end{figure}
In this subsection we add the complication of the magnetic field direction $\bc$ at an arbitrary angle to the toroidal direction. Figure~\ref{fig:rectInboardArbitraryAngle} shows the additional angles $\alphac$ and $\betac$ required to specify this angle. The `polar' angle $\betac$ varies from 0 in the toroidal direction (pointing counterclockwise as seen from above) to $\pi$ in the anti-toroidal (clockwise) direction.
The `azimuthal' angle $\alphac$ determines the plane in which $\betac$ lies. It is importantly only when $\betac$ is different from 0 or $\pi$. It rotates counterclockwise around the positive \textit{toroidal} direction, so $\alpha=0$ is radially outward, $\alpha=\pi/2$ is $-\hat{z}$ and so on.

The magnetic field is axisymmetric. Therefore the value of $\alphac$ and $\betac$ are independent of $\phi$, and $\bc$ rotates around the $\hat{z}$ axis as $\phi$ is varied.
In Cartesian coordinates,
\begin{equation}
\label{eq:bcangle}
\bc = (\cos \alphac \sin\betac \cos\phi-\cos\betac \sin\phi,\quad\cos\betac \cos\phi+\cos\alphac \sin\betac \sin\phi,\quad-\sin\alphac \sin\betac),
\end{equation}
and
\begin{equation}
\begin{split}
\cos\thetac &= \frac{\bc \cdot \Deltac}{|\Deltac|} \\
  &=\frac{(\pc-\uc \cos \phi ) \cos\alphac\sin \betac +\uc \cos \betac  \sin \phi -\zc \sin \alphac  \sin \betac }{\sqrt{\pc^2 + \uc^2+\zc^2-2 \pc \uc \cos \phi}}.
\end{split}
  \label{eq:costhetacangle}
\end{equation}
In this section the numerators of the integrand are lengthy, but the denominators are all the same, with the value $|\Deltac|^5 = \left(\pc^2 + \uc^2+\zc^2-2 \pc \uc \cos \phi\right)^{5/2}$.
The integrands can then be decomposed into sums of terms with $\alphac$ and $\betac$, $\pc$, $\uc$, and $\zc$, which are all independent of $\phi$, multiplied by `components' $\cos^m\phi\sin^{n}\phi/|\Deltac|^{5/2}$, where $0 \le \{m, n\} \le 3$ are integers.
Section~\ref{sec:compint} lists these component integrals.

\begin{landscape}
\subsubsection{Proportional to \texorpdfstring{$\sin^2\thetac$}{Sine theta squared}, A mode}
\begin{equation}
\begin{split}
  g_{RiAa} = \int_{-\phi_i}^{\phi_i}\,d\phi \frac{\pc \cos\nuc \sin^2\thetac}{|\Deltac|^2} &= \int_{-\phi_i}^{\phi_i} \frac{d\phi}{|\Deltac|^{5}}\Big[
  \left(-u p \left(u^2+p^2+z^2\right)+u p (p \cos \alpha -z \sin \alpha )^2 \sin ^2\beta \right) * 1 \\
  &+ \Bigl(3 u^2 p^2+p^4+p^2 z^2-p^2 \left(2 u^2+p^2\right) \cos ^2\alpha \sin ^2\beta-p^2 z^2 \sin ^2\alpha \sin ^2\beta %\\
  +u^2 z p \sin (2 \alpha ) \sin ^2\beta +p^3 z \sin 2 \alpha \sin ^2\beta\Bigr) * \cos\phi \\
  &+u p \left(\left(u^2+2 p^2\right) \cos ^2\alpha  \sin ^2\beta -p \left(2 p+z \sin (2 \alpha) \sin ^2\beta \right)\right) * \cos^2\phi \\
  & -u^2 p^2 \cos ^2\alpha  \sin ^2\beta * \cos^3\phi % \\
    + u^3 p \cos ^2\beta * \sin^2\phi %\\
   -u^2 p^2 \cos ^2\beta * \cos\phi \sin^2\phi\Big].
\end{split}
\label{eq:gRiAasetup}
\end{equation}
The numerator contains some terms with $\sin\phi$, $\cos\phi \sin\phi$, and $\cos^2\phi \sin\phi$; these are zero by symmetry.
The integral evaluates to
\begin{equation}
    \begin{split}
        g_{RiAa} &= -\frac{2 \sqrt{p^2-u^2}}{3 p q^2 \sqrt{p^2-u^2+z^2}}\biggl\{
              2 q \left(-2 p^4+p^2 (2 u^2-z^2)+z^2 (u^2+z^2)\right)
            + \biggl[-3 q z^2 (p^2+u^2+z^2) \\
            &+ \Bigl(2 p^8+p^6 \left(z^2-6 u^2\right)+p^4 \left(6 u^4-11 u^2 z^2-5 z^4\right)+p^2 \left(-2 u^6+11 u^4 z^2+8 u^2 z^4-5 z^6\right)-z^2 \left(u^2+z^2\right)^3 \Bigr)\cos(2\alphac) \\
            &-p z \Bigl(5 p^6+p^4 \left(11 z^2-7 u^2\right)-p^2 \left(u^4+18 u^2 z^2-7 z^4\right)+\left(u^2+z^2\right)^2 \left(3 u^2+z^2\right)\Bigr)\sin(2\alphac)
            \biggr]\sin^2(\betac)
         \biggr\} \\
        &+\frac{F}{3 p q u \sqrt{(p-u)^2+z^2}}\biggl\{
            2 q (-p^2+u^2+2 z^2)+\Bigl[-3 q (p^2+u^2+2 z^2)\\
        & +2 \left((p^2-u^2)^3-2 (p^4-3 p^2 u^2+2 u^4) z^2-5 (p^2+u^2) z^4-2 z^6\right) \cos (2 \alphac ) \\
        & +2 p z \left(-2 (p^2-u^2)^2-3 (p^2+u^2) z^2-z^4\right) \sin (2 \alphac )\Bigr] \sin ^2(\betac )
        \biggr\} \\
        &+\frac{\sqrt{(p-u)^2+z^2}\;E}{3 p q^2 u}\biggl\{
            -2 q \Bigl(-p^4+u^4+(p^2+3 u^2) z^2+2 z^4\Bigr)+\biggl[3 q \left((p^2-u^2)^2+3 (p^2+u^2) z^2+2 z^4\right) \\
        & +\left(-p^8+p^6 (2 u^2+z^2)+(u^2+z^2)^3 (u^2+2 z^2)-p^2 (u^2+z^2) (2 u^4+15 u^2 z^2-7 z^4)+p^4 (11 u^2 z^2+7 z^4)\right) \cos (2 \alphac ) \\
        & +2 p z \left(2 (p^2-u^2)^2 (p^2+u^2)+5 (p^2-u^2)^2 z^2+4 (p^2+u^2) z^4+z^6\right) \sin (2 \alphac )\biggr] \sin ^2(\betac ) \biggr\}
    \end{split}
    \label{eq:gRiAa}
\end{equation}
where incomplete elliptic integrals $F(\ellamp, \ellpar)$ and $E(\ellamp, \ellpar)$ have amplitude $\ellamp = \phi_i/2 = \cos^{-1}(u/p)/2$ and parameter $\ellpar = -4 p u/((p - u)^2 + z^2))$,
and where the factor $q\equiv p^4-2 p^2 (u^2-z^2)+(u^2+z^2)^2$.

\subsubsection{Proportional to \texorpdfstring{$\sfrac{1}{4} + \sfrac{3}{4}\cos^2\thetac$}{1/4 + 3/4 cosine theta squared}, B/C modes}
The B and C modes exhibit emission proportional to $\sfrac{1}{4} + \sfrac{3}{4}\cos^2\thetac$.
Combining $\sfrac{1}{4}\,g_{RiI}$$ + \sfrac{3}{4}\,g_{Rica}$ (Eqs.~\ref{eq:gRiI} and~\ref{eq:gRica}, resp.) yields
  \begin{equation}
    \begin{split}
      g_{RiBa} = & \frac{\sqrt{p^2-u^2}}{2 p q^2 \sqrt{p^2-u^2+z^2}} \biggl\{
        2 q \left(2 p^2 (p^2-u^2) + (3 p^2+u^2) z^2+z^4\right)+\biggl[-3 q z^2 (p^2+u^2+z^2) \\
        & +\left(2 p^8+p^6 (-6 u^2+z^2)-z^2 (u^2+z^2)^3+p^4 (6 u^4-11 u^2 z^2-5 z^4)+p^2 (-2 u^6+11 u^4 z^2+8 u^2 z^4-5 z^6)\right) \cos (2 \alphac )\\
        &-p z \left(5 p^6+(u^2+z^2)^2 (3 u^2+z^2)+p^4 (-7 u^2+11 z^2)-p^2 (u^4+18 u^2 z^2-7 z^4)\right) \sin (2 \alphac )\biggr] \sin ^2(\betac )
      \biggr\} \\
      &+\frac{F}{4 p q u \sqrt{(p-u)^2+z^2}}\biggl\{-2 q (3 p^2+u^2+2 z^2)+\biggl[3 q (p^2+u^2+2 z^2)\\
      &+\left(-(p^2-u^2)^3+2 (p^4-3 p^2 u^2+2 u^4) z^2+5 (p^2+u^2) z^4+2 z^6\right) \cos (2 \alphac ) \\
      &+2 p z \left(2 (p^2-u^2)^2+3 (p^2+u^2) z^2+z^4\right) \sin (2 \alphac )\biggr] \sin ^2(\betac )\biggr\} \\
      &+\frac{\sqrt{(p-u)^2+z^2}\,E}{4 p q^2 u}\biggl\{
2 q \left(3 p^4+u^4+3 u^2 z^2+2 z^4+p^2 (-4 u^2+5 z^2)\right)+\biggl[-3 q \left((p^2-u^2)^2+3
(p^2+u^2) z^2+2 z^4\right)\\
      &+\left(p^8-p^6 (2 u^2+z^2)-(u^2+z^2)^3 (u^2+2 z^2)+p^2 (u^2+z^2) (2 u^4+15 u^2 z^2-7 z^4)-p^4 (11 u^2 z^2+7 z^4)\right) \cos (2 \alphac )\\
      &+2 p z \left(-2 (p^2-u^2)^2 (p^2+u^2)-5 (p^2-u^2)^2 z^2-4 (p^2+u^2) z^4-z^6\right) \sin (2 \alphac )\biggr] \sin ^2(\betac ) \biggr\}
    \end{split}
    \label{eq:gRiBa}
  \end{equation}
with $F$ and $E$ as for $\sin^2\thetac$.
This ends the section on the intensity on a vertical inboard wall (center stack) of an axisymmetric torus.
\end{landscape}

\subsection{Horizontal floor or ceiling of a rectangular torus}
\label{ssec:floor}
\begin{figure}[htb]
    \centering
    \includegraphics[width=0.75\textwidth]{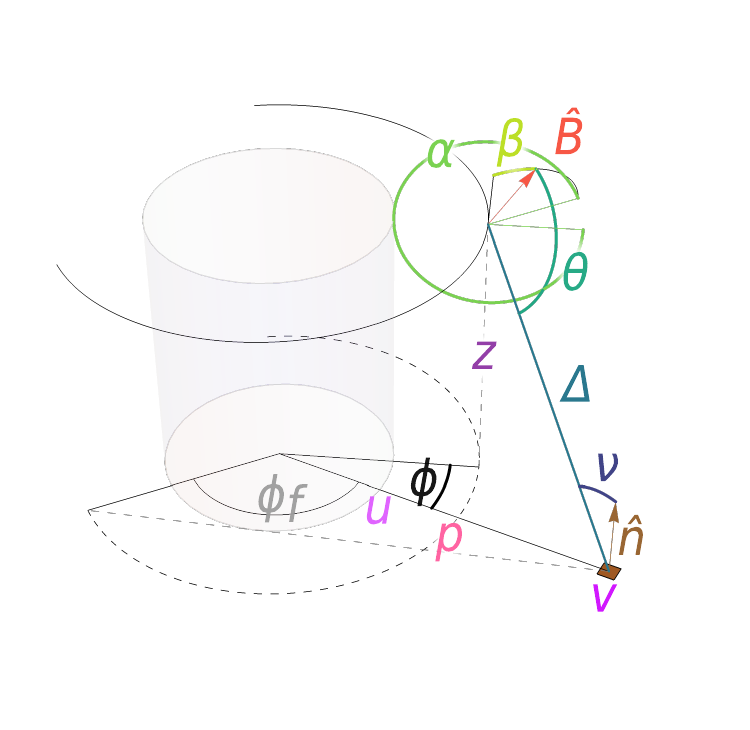}
    \caption{Geometry to compute the intensity on a target patch of the floor at radius $\vc$, from an infinitesimal ring source of plasma at radius $\pc$ and relative height $\zc$, and where
    the inboard wall has radius $\uc$.
    The visible range of the ring source $\pm \phi_f$ depends on $\vc$ as well as $\uc$ and $\pc$.
    See earlier figures for explanations of $\alphac$, $\betac$, $\thetac$, $\Deltac$, $\nuc$, and $\normc$.
    }
    \label{fig:rectFloor}
\end{figure}
The calculation of intensities on the floor proceed similar to those in Subsection~\ref{ssec:inboardToroidalField} and~\ref{ssec:inboardAngledField}.
The differences will be noted here.
First, the target location is $T = (\vc, 0, 0)$ and the normal is $\normc = (0,0,1)$.
The difference vector is $\Deltac = (p \cos \phi -v , p \sin \phi , z)$, and therefore,
$$\cos\nuc = \frac{\normc\cdot\Deltac}{|\Deltac|} = \frac{\zc}{|\Deltac|} = \frac{\zc}{\sqrt{\pc^2+\vc^2+\zc^2-2 \pc \vc \cos\phi}}.$$
The range of integration over $\phi$ is
$$\pm \phi_f \equiv \cos^{-1}\bigg(\frac{\uc}{\pc}\bigg)+ \cos^{-1}\bigg(\frac{\uc}{\vc}\bigg).$$
The expression for $\bc(\alphac,\betac, \phi)$ is the same as previously in Eq.~\ref{eq:bcangle}.
The expression for $\cos\thetac$ is the same as Eq.~\ref{eq:costhetacangle} except $\vc$ takes the place of $\uc$:
\begin{equation}
\begin{split}
\cos\thetac &= \frac{\bc \cdot \Deltac}{|\Deltac|} \\
  &=\frac{(\pc-\vc \cos \phi ) \cos\alphac\sin \betac +\vc \cos \betac  \sin \phi -\zc \sin \alphac  \sin \betac }{\sqrt{\pc^2 + \vc^2+\zc^2-2 \pc \vc \cos \phi}}.
  \label{eq:floorcosthetacangle}
\end{split}
\end{equation}

\subsubsection{Isotropic}
The reduced intensity on a horizontal patch is
\begin{equation}
  g_{RfI} = \int_{\phi_f}^{\phi_f}\,d\phi \frac{\pc \cos\nuc}{|\Deltac|^2} = \int_{\phi_f}^{\phi_f}\,d\phi\,\frac{p z}{\left(p^2+v^2+z^2-2 p v \cos (\phi )\right)^{3/2}}.
  \label{eq:gRfIsetup}
\end{equation}
This evaluates to
\begin{equation}
  \begin{split}
    g_{RfI} &= \frac{8 p u \left(\sqrt{p^2-u^2}+\sqrt{v^2-u^2}\right) z}{q\,\sqrt{p^2 + v^2 -2 u^2 + 2 \sqrt{(p^2-u^2)(v^2-u^2)}+z^2} } \\
            &+ \frac{4 p z\sqrt{(p-v)^2+z^2} E(\ellamp, \ellpar)}{ q \left((p+v)^2+z^2\right)}
  \end{split}
    \label{eq:gRfI}
\end{equation}
where here in Section~\ref{ssec:floor}, $q\equiv p^4-2 p^2 \left(v^2-z^2\right)+\left(v^2+z^2\right)^2$, the incomplete elliptical integral $E(\ellamp, \ellpar)$ has amplitude $\ellamp = \phi_f/2 = \left(\cos^{-1}\left(\uc/\pc\right) + \cos^{-1}\left(\uc/\vc\right)\right)/2$
and parameter $\ellpar = -4 p v/((p-v)^2+z^2)$. While this formula has no $F(\ellamp, \ellpar)$ term, subsequent ones in this subsection do and its $\ellpar$ and $\ellamp$ are the same.

\begin{landscape}
\subsubsection{Proportional to \texorpdfstring{$\sin^2\thetac$}{Sine theta squared}, A mode}
We present the reduced intensities from polarized sources in an angled field. For $\sin^2\thetac$,
\begin{equation}
\begin{split}
    g_{RfAa} = \int_{-\phi_f}^{\phi_f}\,d\phi \frac{\pc \cos\nuc \sin^2\thetac}{|\Deltac|^2} = \int_{-\phi_f}^{\phi_f} \frac{d\phi}{|\Deltac|^{5}}\Big[&
 p z \left(p^2+v^2+z^2-(p \cos \alpha -z \sin \alpha )^2 \sin ^2\beta \right) * 1
  + 2 p v z \left(\cos \alpha  (p \cos \alpha -z \sin \alpha ) \sin ^2\beta  - p\right)* \cos\phi \\
  &-p v^2 z \cos ^2\alpha  \sin ^2\beta * \cos^2\phi -p v^2 z \cos ^2\beta * \sin^2\phi \Big].
\end{split}
\end{equation}
With $t\equiv\sqrt{(p^2-u^2)(v^2-u^2)}$
this evaluates to
\begin{equation}
    \begin{split}
        & g_{RfAa} =
          \frac{u \left(\sqrt{p^2-u^2}+\sqrt{v^2-u^2}\right) z}{3 p \left(p^2+2 t-2 u^2+v^2+z^2\right)^{3/2}\,q^2} \biggl\{ 
           2 q \left(11 p^4-\left(v^2+z^2\right) \left(4 t-4 u^2+v^2+z^2\right)+2 p^2 \left(10 t-10 u^2+3 v^2+5 z^2\right)\right)\\
           &+\biggl[-3 q \left(3 p^4+2 p^2 \left(2 t-2 u^2-v^2+z^2\right)-\left(v^2+z^2\right) \left(4 t-4 u^2+v^2+z^2\right)\right)
            +\Bigl(-11 p^8+\left(v^2+z^2\right)^3 \left(4 t-4 u^2+v^2+z^2\right)+4 p^6 \left(-5 t+5 u^2+4 v^2+2 z^2\right)\\
            &-4 p^2 \left(v^2+z^2\right) \left(v^2 \left(7 t-7 u^2+2 v^2\right)+\left(-13 t+13 u^2-6 v^2\right) z^2-8 z^4\right)+2 p^4 \left(22 \left(t-u^2\right) v^2+v^4+2 \left(7 t-7 u^2+9 v^2\right) z^2+25 z^4\right)\Bigr) \cos (2 \alphac )\\
           &+8 p z \left(4 p^6-2 p^2 (v^2-z^2) \left(3 t-3 u^2+v^2+z^2\right)-\left(v^2+z^2\right)^2 \left(t-u^2+v^2+z^2\right)+p^4 \left(7 t-7 u^2-v^2+7 z^2\right)\right) \sin (2 \alphac )\biggr] \sin ^2(\betac ) \biggr\} \\
      &+\frac{z\;F}{3 p q \sqrt{p^2-2 p v+v^2+z^2}}\biggl\{
    2 q-\left(3 q+\left(p^4-2 p^2 \left(v^2-3 z^2\right)+\left(v^2+z^2\right)^2\right) \cos (2 \alphac )+2 p z \left(p^2-v^2-z^2\right) \sin (2 \alphac )\right) \sin ^2(\betac ) \biggr\} \\
  &-\frac{z \sqrt{(p-v)^2+z^2}\;E}{3 p q^2}\biggl\{2 q \left(5 p^2-v^2-z^2\right)+\biggl[3 q \left(-p^2+v^2+z^2\right)+\left(-5 p^6+\left(v^2+z^2\right)^3+p^4 \left(11 v^2+7 z^2\right)+p^2 \left(-7 v^4+6 v^2 z^2+13 z^4\right)\right) \cos (2 \alphac ) \\
      &+2 p z \left(7 p^4-6 p^2 \left(v^2-z^2\right)-\left(v^2+z^2\right)^2\right) \sin (2 \alphac )\biggr] \sin ^2(\betac) \biggr\}.
\end{split}
\label{eq:RfAa}
\end{equation}
\newpage

\newpage
\subsubsection{Proportional to \texorpdfstring{$\sfrac{1}{4} + \sfrac{3}{4}\cos^2\thetac$}{1/4 + 3/4 cosine theta squared}, B/C modes}
Combining $\sfrac{1}{4}\;g_{RfI} + \sfrac{3}{4}\;g_{RfCa}$ (Eqs.~\ref{eq:gRfI} and~\ref{eq:gRfCa}, resp.) yields
\begin{equation}
    \begin{split}
      &g_{RfBa}=\frac{u \left(\sqrt{p^2-u^2}+\sqrt{-u^2+v^2}\right) z}{4 p q^2 (p^2+2 t-2 u^2+v^2+z^2)^{3/2}}\biggl\{
2 q \left(5 p^4+(v^2+z^2) (4 t-4 u^2+v^2+z^2)+2 p^2 (6 t-6 u^2+5 v^2+3 z^2)\right)\\
      &+\biggl[3 q \left(3 p^4+2 p^2 (2 t-2 u^2-v^2+z^2)-(v^2+z^2) (4 t-4 u^2+v^2+z^2)\right)
       +\Bigl(11 p^8+4 p^6 (5 t-5 u^2-4 v^2-2 z^2)-(v^2+z^2)^3 (4 t-4 u^2+v^2+z^2)\\
       &-2 p^4 \left(v^4+14 t z^2+25 z^4-2 u^2 (11 v^2+7 z^2)+2 v^2 (11 t+9 z^2)\right)+4 p^2 (v^2+z^2) \left(2 v^4-13 t z^2-8 z^4+v^2 (7 t-6 z^2)+u^2 (-7 v^2+13 z^2)\right)\Bigr) \cos (2 \alphac )\\
      &+8 p z \left(-4 p^6+2 p^2 (v^2-z^2) (3 t-3 u^2+v^2+z^2)+(v^2+z^2)^2 (t-u^2+v^2+z^2)+p^4 \left(7 u^2+v^2-7 (t+z^2)\right)\right) \sin (2 \alphac )\biggr] \sin ^2(\betac )
        \biggr\} \\
      &+\frac{z\;F}{4 p q \sqrt{(p-v)^2+z^2}}\biggl\{-2 q+\left(3 q+\left(p^4-2 p^2 (v^2-3 z^2)+(v^2+z^2)^2\right) \cos (2 \alphac )+2 p z (p^2-v^2-z^2) \sin (2 \alphac)\right) \sin ^2(\betac ) \biggr\} \\
      &+\frac{z \sqrt{(p-v)^2+z^2}\;E}{4 p q^2}\biggl\{2 q (3 p^2+v^2+z^2)+\biggl[3 q (p^2-v^2-z^2)+\\
      &\left(5 p^6+p^2 (7 v^2-13 z^2) (v^2+z^2)-(v^2+z^2)^3-p^4 (11 v^2+7 z^2)\right) \cos (2 \alphac )
       +2 p z \left(-7 p^4+6 p^2 (v^2-z^2)+(v^2+z^2)^2\right) \sin (2 \alphac )\biggr] \sin ^2(\betac )
        \biggr\}.
    \end{split}
  \label{eq:gRfBa}
\end{equation}
\end{landscape}

\subsection{Outboard wall}\label{ssec:outerwall}
\begin{figure}[h]
    \centering
    \includegraphics[width=0.75\textwidth]{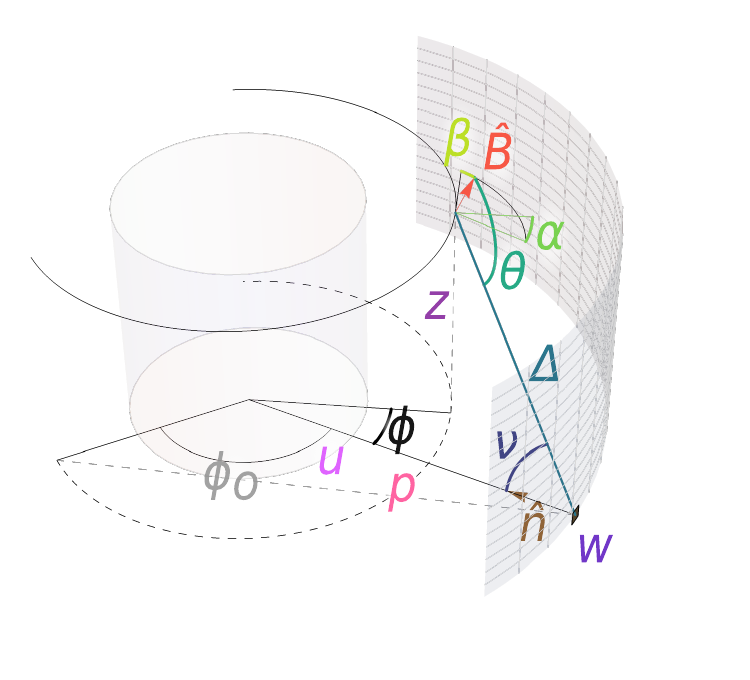}
    \caption{Geometry for the intensity on the outboard wall, which has radius $\wc$.
    }
    \label{fig:rectOutboard}
\end{figure}
The calculations for reduced intensity at the outboard are trivially relatable to the ones for the inboard wall.
Figure~\ref{fig:rectOutboard} shows the geometry; it names $\wc$ as the radius of the outer wall.
The target location is $T=(\wc,0,0)$ and its normal is $\normc=(-1,0,0)$.
The range of integration over $\phi$ is
$$\pm \phi_o \equiv \cos^{-1}\biggl(\frac{\uc}{\pc}\biggr) + \cos^{-1}\biggl(\frac{\uc}{\wc}\biggr).$$
To calculate the reduced intensity, use the arbitrary-integration-angle formulas for the \textit{inboard} side with three modifications.
\begin{enumerate}
\item The expression is negated, due to the normal in the opposite direction.
\item Use $\wc$ in place of $\uc$. Since $\wc > \pc$ while $\uc < \pc$ this ensures a positive reduced intensity.
\item Use $\phi_o$ in place of $\phi_i$.
\end{enumerate}

With the formulas presented in Subsections~\ref{ssec:inboardAngledField}--\ref{ssec:outerwall}, we can calculate the NWL on the interior of a torus with a rectangular cross section such as the one presented above in the example of Section~\ref{sec:rectexample}.
The next section will develop the formulas required for a torus with an arbitrary convex cross section, and examples employing these formulas will be shown in Section~\ref{sec:uniformNWLexamples}.

% The difference vector $\Deltac = (\pc \cos\phi - \wc, \pc\sin\phi, \zc)$ and therefore
% $$ \cos\nuc = \frac{\normc\cdot\Deltac}{|\Deltac|} = \frac{\wc-\pc\cos\phi}{\sqrt{\pc^2+\wc^2+\zc^2 - 2 \pc \wc \cos\phi}}.$$
% The cosine of $\thetac$ is
% $$\cos\thetac = \frac{\bc\cdot\Deltac}{|\Deltac|} = \frac{(\pc-\wc \cos \phi )\cos \alphac \sin \betac +\wc \cos \betac  \sin \phi -\zc \sin \alphac \sin \betac }{\sqrt{\pc^2+\wc^2+\zc^2 - 2 \pc \wc \cos\phi}}.$$

% \subsubsection{Isotropic}
% The reduced intensity on a patch on the outer wall is
% \begin{equation}
%     g_{RoI} = \int_{\phi_o}^{\phi_0}\,d\phi \frac{\pc \cos\nuc}{|\Deltac|^2} = \int_{\phi_o}^{\phi_0}\,d\phi\,\frac{p (w-p \cos \phi )}{\left(p^2+w^2+z^2-2 p w \cos \phi \right)^{3/2}}.
% \end{equation}
% This evalutes to $g_{RoI} &= -g_{RiI\phi}(\phi_m \to \phi_o, u\to w)$

\section{Interior of torus with arbitrary convex cross section}
\label{sec:arbtorus}
We can calculate the neutron wall load (NWL) for a torus with an \textit{arbitrary} convex cross section by using variants of the formulas already presented for the rectangular-cross-section torus.
There are two changes that need to be made.
First, we need to calculate the integration bounds $\pm\phi_m$ for the NWL on an element from each ring of plasma.
% We will present the algorithm used for a \textit{convex} cross section, but suggest how it can be generalized to an arbitrary cross section.
The algorithm will be presented.
Second, the NWL on a target wall element tilted at an arbitrary angle can be computed with a weighted sum of the formulas for horizontal and vertical targets with those integration bounds.
We will show how the leading terms and the $F$ and $E$ terms can be altered from the $Ri$ and $Rf$ cases into the general horizontal-normal-component $H$ and vertical-normal-component $V$ forms.

\subsection{Integration bounds}
\label{ssec:arbtorusbounds}
\begin{figure}[htb]
    \centering
    \includegraphics[width=0.95\textwidth]{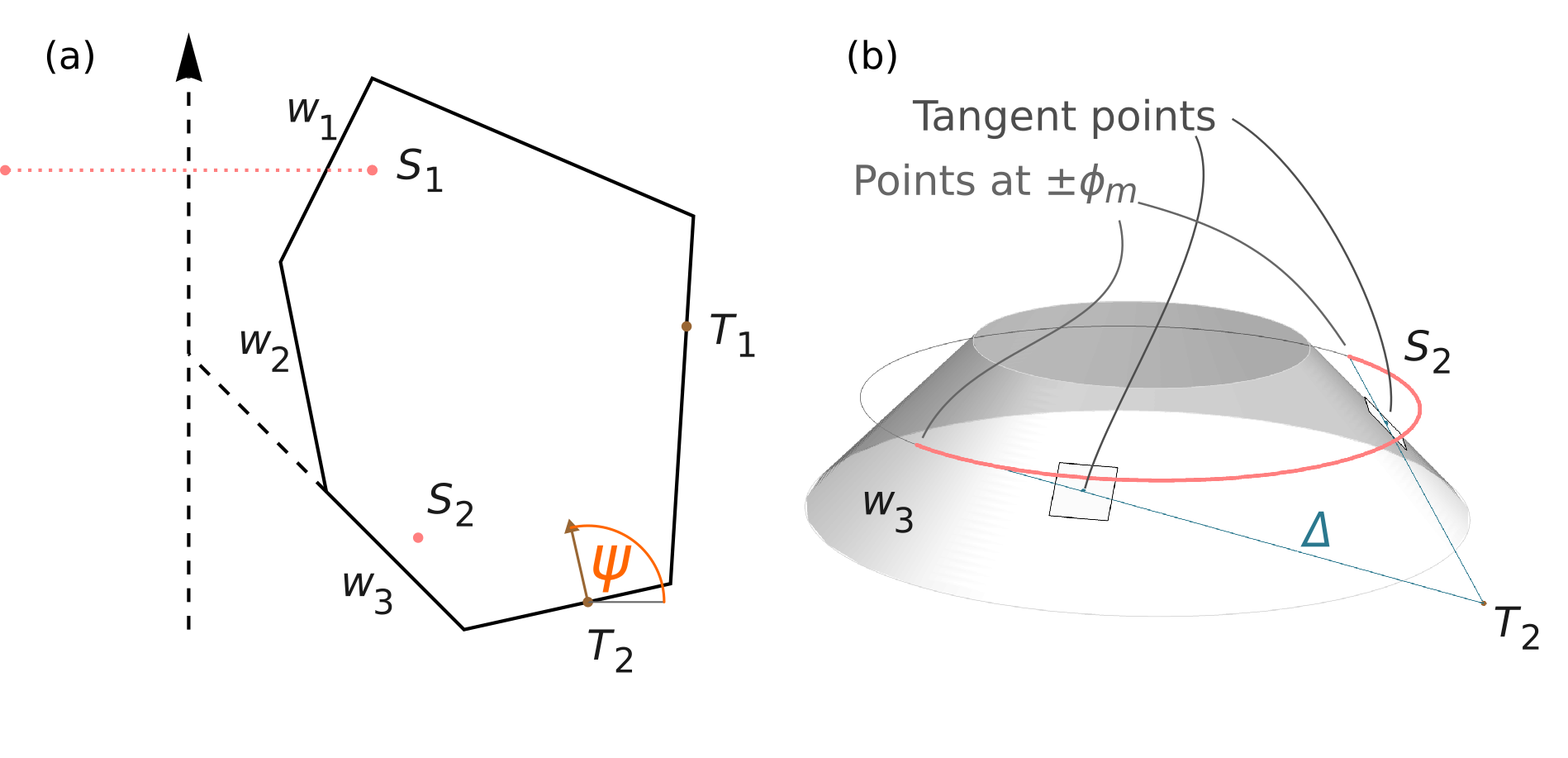}
    \caption{Part (a) shows a cut through a torus with a convex cross section. The three outward-facing wall segments are $w_{1\ldots 3}$. Two source rings are $S_1$, $S_2$. Two targets are $T_1$, $T_2$. The angle between a target's normal and a radial vector is $\psic$.
    Part (b) shows how the visible part of the ring source $S_2$, as seen from $T_2$, is limited to $\pm \phi_m$ where the sightline $\Deltac$ is tangent to the wall segment $w_3$. The tangent points are highlighted with white squares lying on the conical wall segment.}
    \label{fig:convextorus}
\end{figure}

Figure~\ref{fig:convextorus} (a) depicts a cut through a torus with a convex cross section with some example source ring $S$ and target $T$ locations.
The three outward-facing wall segments are labeled $w_{1\ldots 3}$. 
An outward-facing wall segment has a normal pointing in the direction $-\pi/2 < \psic < \pi/2$.
Outward-facing wall segments are the only ones which are important for the integration bounds calculation, as the \textit{inward}-facing wall segments cannot block the line of sight from a target to a source ring.
% redo this sentence
With a convex-cross-section geometry, every source ring is visible from any point on the wall 
and the furthest visible angles $\pm \phi_m$ are always limited by sightlines which are tangent to an outward-facing wall segment.
One example of this is shown in part~(b), where the sightline vectors at $\Deltac = S_2(\pm\phi_m) - T_2$ are tangent to the wall segment $w_3$.

The segments spanning the height range between the source and the target need to be checked to find which yields the most restrictive maximum angle.
For example, between $S_1$ and $T_1$ only $w_1$ and $w_2$ must be checked, between $S_1$ and $T_2$ all three segments must be checked, and between $S_2$ and $T_2$ only $w_3$ must be checked.
 
The source ring angle that satisfies the tangency condition is
\begin{equation}
\phi_m = \arccos\left(\frac{x_p x_r-\sqrt{\left(p^2-x_p^2\right)\left(r^2-x_r^2\right)}}{p\,r}\right)
\label{eq:tangencyU}
\end{equation}
where $p$ and $r$ are the radii of the source and target, respectively, and $x_p$ and $x_r$ are the radii of the wall segment at the heights of the source and target, respectively.
The wall segment being tested may not be present at these heights.
In that case $x_p$ and $x_r$ are calculated by extending the segment, even to $x < 0$.
The parameter
\begin{equation}
t_c \equiv \frac{r^2-x_r^2-\sqrt{\left(p^2-x_p^2\right)\left(r^2-x_r^2\right) }}{(r^2-x_r^2)-(p^2-x_p^2)}
\label{eq:tangencyTc}
\end{equation}
calculates the fraction of the distance from the target to the source where the contact occurs.
The height of the contact point, relative to the target, is $z_c \equiv t_c z$ where $z$ is the height of the source relative to the target, and the contact radius is $r_c \equiv x_r + z_c m$ where $m$ is the inverse slope of the wall segment (a cylindrical, vertical wall segment would have $m=0$).

These formulas can fail to produce valid quantities in several ways, which indicate that a segment is not restricting the visible ring angle.
First, if the discriminant is negative this indicates that the (extended) segment never blocks the view between the source and target.
This is the case for $S_1$ as viewed from $T_2$ when testing $w_3$: all of the ring is above even the extended cone of $w_3$
\footnote{A negative discriminant is also possible if the source and target are on opposite sides of the extended segment, though this cannot occur for convex cross sections.}.
Second, $r_c < 0$ also indicates that the segment never blocks the view.
Finally, if the segment does not include the relative height $z_c$ then it does not block the view.

For a target on an outward-facing wall segment, the formulas for testing that segment reduce to $\phi_m = \arccos(x_p/p)$ and $t_c=0$: the limiting angle yields $\Deltac$ tangent to the surface at the target point itself.

% \subsection{For non-convex cross sections}

% Views can be limited by the endpoints of segments, rather than tangency.

\subsection{Horizontal and vertical components of the normal over arbitrary integration angle}
This subsection will present formulas to compute the neutron wall load (NWL) on a wall element with an arbitrary normal vector poloidal angle $\psic$.
In the above derivations for the reduced intensity on walls with horizontal $(1,0,0)$ or vertical $(0,0,1)$ normals, the integrand has a multiplicative term $\cos\nuc = \normc \cdot \Deltac / |\Deltac| $.
The normal now has components $\normc = (\cos\psic, 0, \sin\psic)$.
The integrand, and the result, can therefore naturally be split into combinations of the horizontal and vertical cases.
As an example, these would be combined like
\begin{equation}\label{eq:emissionsum}
g_\mathrm{total} = (h_{HIL} + h_{HIF} + h_{HIE})\cos\psic + (h_{VIL} + h_{VIE})\sin\psic
\end{equation}
for the isotropic emission case. 
The term $h_{HIL}$ is given in Eq.~\ref{eq:hHIL}; $h_{VIL}$ is presented below.
This section only prints full forms for the leading terms ($L$) because the $F$ and $E$ terms can be related via a simple transformation, as explained just after Eq.~\ref{eq:hHIL}.

\subsubsection{Isotropic distribution, vertical component of normal}
The leading term of $g_{RfI}$, $h_{RfIL}$, becomes
\[
h_{VIL} = \frac{8 p^2 r z \sin \phi_m}{q_r \sqrt{p^2+r^2+z^2-2 p r \cos \phi_m}}
\label{eq:hVIL}
\]
where $q_r\equiv p^4-2 p^2 \left(r^2-z^2\right)+\left(r^2+z^2\right)^2$.
Adapting the $E$ term $h_{RfIE}$ into $h_{VIE}$, for arbitrary $\phi_m$ and target radius, only requires changing $\ellamp = \phi_m/2$ and $v\to r$.
This allows computation of the vertical component of the reduced intensity $g_{VI} = h_{VIL} + h_{VIE}$ for a surface tilted at any angle.

% \begin{equation}
%   \begin{split}
%       h_{HAL} &= \frac{-2\sin(\phi_m)}{3 q \left(p^2+u^2+z^2-2 p u \cos (\phi_m)\right)^{3/2}}
%                 \Bigl(-4 p^6+u^6-6 p^4 z^2+2 z^6+2 u^4 (p^2+2 z^2)\\
%               &\qquad+u^2 (p^2+z^2) (p^2+5 z^2)
%                -p u \left(-7 p^4+u^4-2 p^2 z^2+5 z^4+6 u^2 (p^2+z^2)\right) \cos (\phi_m)\Bigr). \\
%   \end{split}
%   \label{eq:hHAL}
% \end{equation}

% For arbitrary target radius $r$ and integration bounds $\pm\phi_m$ the leading term is
% 
% \begin{equation}
%   \begin{split}
%       h_{HcL} &= \frac{2\sin(\phi_m)}{3 q \left(p^2+r^2+z^2-2 p r \cos (\phi_m)\right)^{3/2}}
%       \Bigl(2 p^6+(r^2+z^2)^2 (r^2+2 z^2)
%                +p^4 (r^2+6 z^2)\\
%               &+p^2 (-4 r^4+6 r^2 z^2+6 z^4)-p r \left(5 p^4-6 p^2 r^2+r^4+2 (5 p^2+3 r^2) z^2+5 z^4\right) \cos (\phi_m)\Bigr).
%   \end{split}
%   \label{eq:hHcL}
% \end{equation}

%For arbitrary target radius $r$ and integration bounds $\pm\phi_m$ the algebraic term is

% \begin{equation}
% \begin{split}
%   h_{HBL} &=\frac{\sin(\phi_m)}{2 q_r \left(p^2+r^2+z^2-2 p r \cos (\phi_m)\right)^{3/2}} 
%     \Bigl(4 p^6+\left(r^2+z^2\right)^2 \left(r^2+2 z^2\right)+p^4 \left(r^2+10 z^2\right)\\
%   &+p^2 \left(-6 r^4+6 r^2 z^2+8 z^4\right)-p r \left(9 p^4+r^4+6 r^2 z^2+5 z^4+2 p^2 \left(-5 r^2+7 z^2\right)\right) \cos (\phi_m)\Bigr).
% \end{split}
%   \label{eq:HBL}
% \end{equation}

%For arbitrary target radius $r$ and arbitrary integration bounds $\pm\phi_m$, the $F$ and $E$ terms only change their amplitude $\ellamp=\phi_m/2$. The leading term must be expressed in a more general form:
\begin{landscape}
\subsubsection{Proportional to \texorpdfstring{$\sin^2\thetac$}{Sine theta squared}, A mode}
The leading term of $g_{RiAa}$ becomes
\begin{equation}
    \begin{split}
        h_{HAaL} &= \frac{\sin (\phi_m)}{3 q_r^2 \left(p^2+r^2+z^2-2 p r \cos (\phi_m)\right)^{3/2}}\biggl\{
            2 q_r \Bigl(4 p^6-p^4 (r^2-6 z^2)-(r^2+z^2)^2 (r^2+2 z^2)-2 p^2 (r^4+3 r^2 z^2)\\
                 &+p r \left(-7 p^4+6 p^2 r^2+r^4-2 \left(p^2-3 r^2\right) z^2+5 z^4\right) \cos (\phi_m)\Bigr)
                  +\biggl[3 q_r \Bigl((-p^2 r+r^3)^2+2 (p^2+r^2) (p^2+2 r^2) z^2+(4 p^2+5 r^2) z^4+2 z^6\\
                 &-p r \left((p^2-r^2)^2+6 (p^2+r^2) z^2+5 z^4\right) \cos (\phi_m)\Bigr)
                  + \Bigl(-4 p^{10}+p^8 (9 r^2-6 z^2)+(r^2+z^2)^4 (r^2+2 z^2)
                  +4 p^2 z^2 (r^2+z^2)^2 (-7 r^2+3 z^2)\\
                 &-4 p^6 (r^4-7 r^2 z^2-2 z^4)+p^4 (-2 r^6+6 r^2 z^4+20 z^6)
                   +p r \Bigl((p^2-r^2)^3 (7 p^2+r^2)+8 (p^6-7 p^4 r^2+7 p^2 r^4-r^6) z^2\\
                 &-2 (5 p^4-22 p^2 r^2+9 r^4) z^4-16 (p^2+r^2) z^6-5 z^8\Bigr) \cos (\phi_m)\Bigl)\cos (2 \alphac )
                  +2 p z \Bigl(5 p^8-4 p^6 (r^2-4 z^2)+(r^2+z^2)^3 (5 r^2+z^2)
                  -2 p^4 (r^4+8 r^2 z^2-9 z^4)\\
                 &-4 p^2 (r^6+4 r^4 z^2+r^2 z^4-2 z^6)-4 p r \left(2 (p^2-r^2)^2 (p^2+r^2)+5 (p^2-r^2)^2 z^2+4 (p^2+r^2) z^4+z^6\right) \cos (\phi_m)\Bigr) \sin (2 \alphac)\biggr] \sin ^2(\betac) \biggr\}.
    \end{split}
    \label{eq:hHAaL}
\end{equation}

For arbitrary target radius $r$ and integration bounds $\pm\phi_m$ the leading term is
\begin{equation}
    \begin{split}
      h_{VAaL} &= \frac{r z \sin(\phi_m)}{3 q_r^2 \left(p^2+r^2+z^2-2 p r \cos (\phi_m)\right)^{3/2}}\biggl\{
        2 q_r \Bigl(11 p^4-(r^2+z^2)^2+2 p^2 (3 r^2+5 z^2)+4 p r (-5 p^2+r^2+z^2) \cos (\phi_m)\Bigr)\\
       &+\biggl[-3 q_r \Bigl(3 p^4+2 p^2 (-r^2+z^2)-(r^2+z^2)^2+4 p r (-p^2+r^2+z^2) \cos (\phi_m)\Bigr) 
         + \Bigl(-11 p^8-8 p^2 (r^2-4 z^2) (r^2+z^2)^2 +(r^2+z^2)^4\\
         &+8 p^6 (2 r^2+z^2)+2 p^4 (r^4+18 r^2 z^2+25 z^4)+4 p r \left(5 p^6+p^2 (7 r^2-13 z^2) (r^2+z^2)-(r^2+z^2)^3-p^4 \left(11 r^2+7 z^2\right)\right) \cos (\phi_m)\Bigr)\cos(2\alphac)\\
         &+8 p z \left(4 p^6-p^4 (r^2-7 z^2)-(r^2+z^2)^3+2 p^2 (-r^4+z^4)+p r \left(-7 p^4+6 p^2 (r^2-z^2) +(r^2+z^2)^2\right) \cos (\phi_m)\right) \sin (2 \alphac)\biggr] \sin ^2(\betac) \biggr\}.
\end{split}
\label{eq:hVAaL}
\end{equation}

\subsubsection{B/C modes}
The leading term of $g_{RiBa}$ becomes
\begin{equation}
    \begin{split}
        h_{HBaL} &= \frac{\sin(\phi_m)}{4 q_r^2 \left(p^2+r^2+z^2-2 p r \cos (\phi_m )\right)^{3/2}}\biggl\{
2 q_r \Bigl(4 p^6+(r^2+z^2)^2 (r^2+2 z^2)+p^4 (r^2+10 z^2)+p^2 (-6 r^4+6 r^2 z^2+8 z^4)\\
      &-p r \left(9 p^4+r^4+6 r^2 z^2+5 z^4+2 p^2 (-5 r^2+7 z^2)\right) \cos (\phi_m )\Bigr)+\biggl[3 q_r
   \Bigl(-p^4 (r^2+2 z^2)-(r^2+z^2)^2 (r^2+2 z^2)\\
      &+2 p^2 (r^4-3 r^2 z^2-2 z^4)+p r \left((p^2-r^2)^2+6 (p^2+r^2) z^2+5 z^4\right) \cos (\phi_m )\Bigr)+ \Bigl(4 p^{10}+4 p^2 z^2 (7 r^2-3 z^2) (r^2+z^2)^2\\
      &-(r^2+z^2)^4 (r^2+2 z^2)+p^8 (-9 r^2+6 z^2)+4 p^6 (r^4-7 r^2 z^2-2 z^4)+2 p^4 (r^6-3 r^2 z^4-10 z^6)\\
      &+p r \left(-7 p^8+4 p^6 (5 r^2-2 z^2)+(r^2+z^2)^3 (r^2+5 z^2)+4 p^2 (r^2+z^2) (r^4-15 r^2 z^2+4 z^4)+2 p^4 (-9 r^4+28 r^2 z^2+5 z^4)\right) \cos (\phi_m )\Bigr)\cos(2\alphac)\\
      &-2 p z \Bigl(5 p^8-4 p^6 (r^2-4 z^2)+(r^2+z^2)^3 (5 r^2+z^2)-2 p^4 (r^4+8 r^2 z^2-9 z^4)-4 p^2 (r^6+4 r^4 z^2+r^2 z^4-2 z^6)\\
      &-4 p r \left(2 (p^2-r^2)^2 (p^2+r^2)+5 (p^2-r^2)^2 z^2+4 (p^2+r^2) z^4+z^6\right) \cos (\phi_m )\Bigr) \sin (2 \alphac )\biggr] \sin ^2(\betac) \biggr\}.
    \end{split}
    \label{eq:hHBaL}
\end{equation}

The corresponding leading term for $g_{RfBa}$ is 
\begin{equation}
  \begin{split}
    h_{VBaL} &=\frac{r z \sin (\phi_m)}{4 q_r^2 \left(p^2+r^2+z^2-2 p r \cos (\phi_m)\right)^{3/2}}\biggl\{
2 q_r \Bigl(5 p^4+(r^2+z^2)^2+2 p^2 (5 r^2+3 z^2)-4 p r (3 p^2+r^2+z^2) \cos (\phi_m)\Bigr)\\
    &+\biggl[3 q_r \Bigl(3 p^4-2 p^2 (r^2-z^2)-(r^2+z^2)^2+4 p r (-p^2+r^2+z^2) \cos (\phi_m)\Bigr)+\Bigl(11 p^8+8 p^2 (r^2-4 z^2) (r^2+z^2)^2-(r^2+z^2)^4\\
    &-8 p^6 (2 r^2+z^2)-2 p^4 (r^4+18 r^2 z^2+25 z^4)+4 p r \left(-5 p^6+(r^2+z^2)^3+p^4 (11 r^2+7 z^2)+p^2 (-7 r^4+6 r^2 z^2+13 z^4)\right) \cos (\phi_m)\Bigr)\cos(2\alphac)\\
    &-8 p z \left(4 p^6-p^4 (r^2-7 z^2)-(r^2+z^2)^3+2 p^2 (-r^4+z^4)+p r \left(-7 p^4+6 p^2 (r^2-z^2) +(r^2+z^2)^2\right) \cos (\phi_m)\right) \sin (2 \alphac )\biggr] \sin ^2(\betac )
      \biggr\}.
  \end{split}
  \label{eq:hVBaL}
\end{equation}

\end{landscape}

\section{\textit{Anarr\'{i}ma}, a python package to compute neutron wall loads}\label{sec:python}
The expressions for the reduced intensities ($g_{\ldots}$) described above and the calculation of the integration bounds for a convex-cross section torus (Sec.~\ref{ssec:arbtorusbounds}) have been implemented in python in a package called \anar{}\footnote{This word means ``Sun-border'' in a fictional language\cite{tolkienEtymologies1987, tolkiengatewayAnarrima}.}.
\Anar{} is fully compatible with the automatic differentiation library \textsc{jax}, allowing both forward- and reverse-mode differentiation and just-in-time compilation for accelerated computation.
As of this writing, the special functions library in \textsc{jax} does not contain implementations of the incomplete elliptic integrals $F$ and $E$ required for the reduced intensity calculations.
\Anar{} contains bespoke \textsc{jax}-compatible implementations of the symmetric forms of the elliptic integrals $R_F$ and $R_D$  following the algorithms by Carlson\cite{carlsonComputingEllipticIntegrals1979,carlsonNumericalComputationReal1995,carlsonThreeImprovementsReduction2002}.
The Legendre forms $F$ and $E$ are then easily related to the Carlson symmetric forms.
The computations for the examples of Sec.~\ref{sec:rectexample} and the next section are performed using this package.

\section{Example: Toroidal surfaces with uniform NWL}\label{sec:uniformNWLexamples}
\subsection{Surfaces around a single source ring with a toroidal field}
\begin{figure}[htb]
    \centering
    \includegraphics[trim={0 1.2cm 0 1.5cm},clip,width=0.99\textwidth]{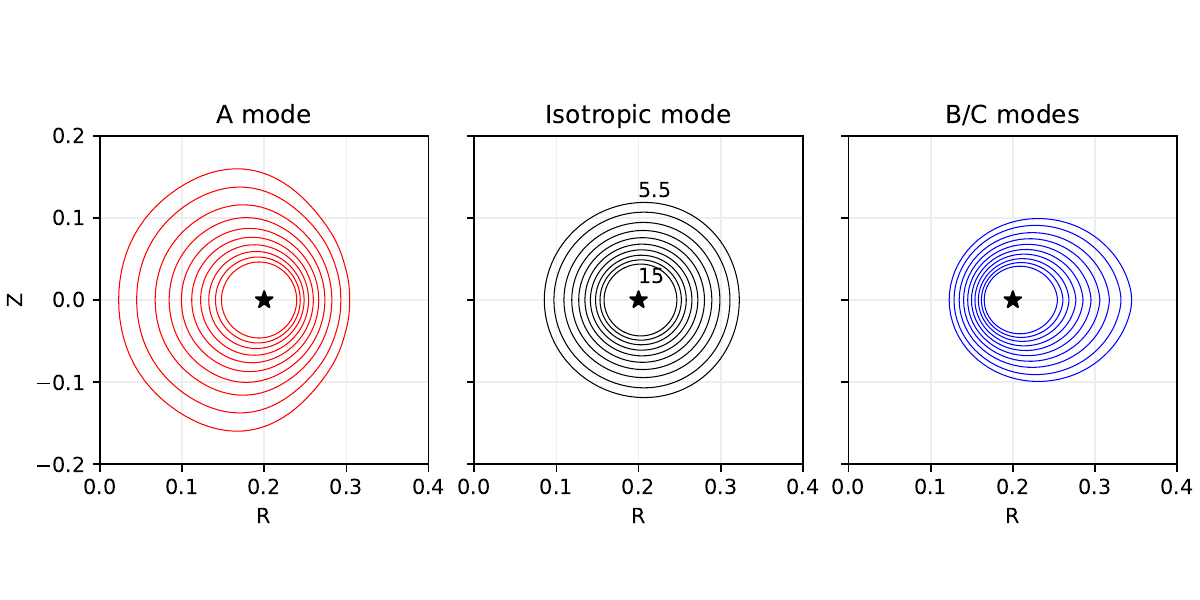}
    \caption{Toroidal shells with uniform NWL from a ring source located at $R=0.2$ units, $\star$, for the three main polarization modes. The ring source is normalized to have a strength of $4\pi/3$ per unit length. In each case the outermost shell has a NWL of 5.5 per unit area and the innermost 15 per unit area; the NWL values of the shells are spaced geometrically.}
    \label{fig:nwl_ring_shells}
\end{figure}

As a demonstration, by using the above formulas we can construct, via a gradient-based optimization scheme, a set of toroidal shells with uniform neutron wall loading (NWL). Figure~\ref{fig:nwl_ring_shells} shows these shells for the A mode, the isotropic mode, and the B/C modes, from a source ring with a purely toroidal field.
In Fig.~\ref{fig:rectanglemak} where plasma was centered in the torus, the A mode had a higher NWL on the inner, top, and bottom walls than the outer wall.
The shape of A mode shells with a uniform NWL compensates for this by moving further away from the source ring on the top, bottom, and inner wall.
Likewise, the uniform-NWL shape for the B/C modes, which send more neutrons outward, bulges on the outside.
Note that these shapes are \textit{not} contours of neutron flux or similar, because in each instance the local neutron flux is determined by the fraction of the source ring which is visible, limited by the walls.

\begin{figure}[htb]
    \centering
    \includegraphics[width=0.99\textwidth]{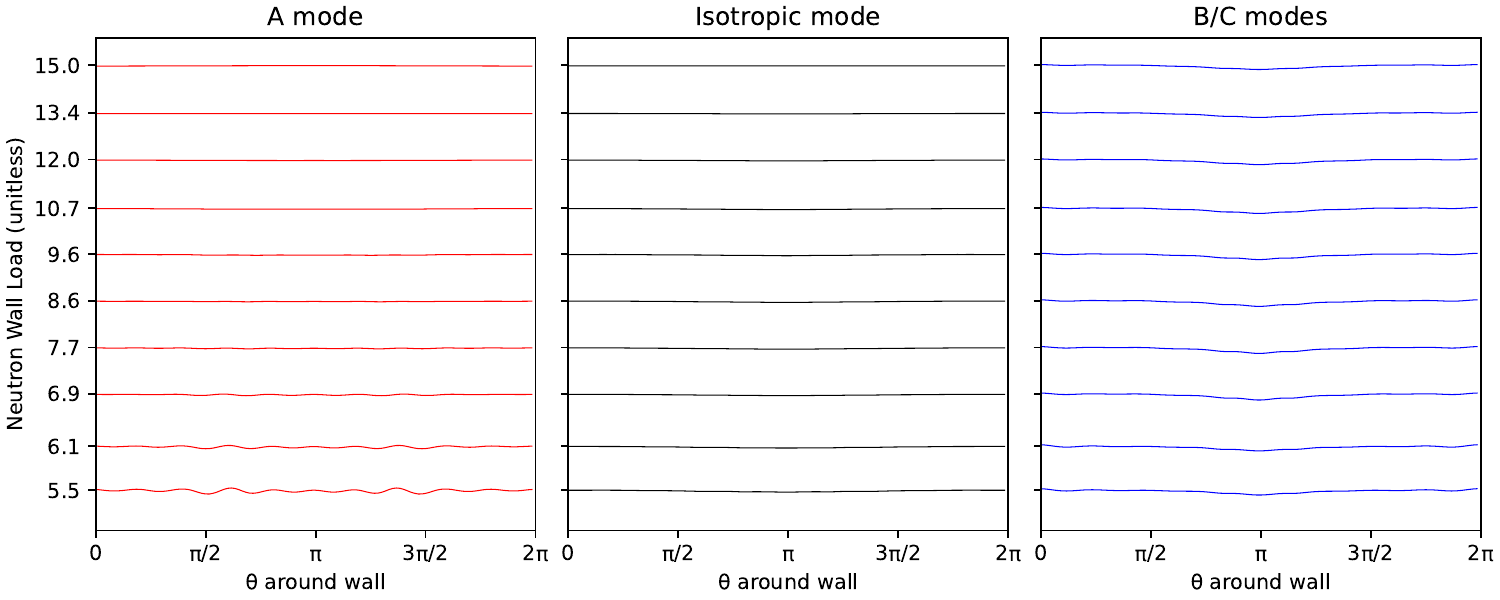}
	\caption{Evaluated NWL on the surfaces shown in Fig.~\ref{fig:nwl_ring_shells}. The residual nonuniformity is is because the wall shape is described using a small number of Fourier modes.}
    \label{fig:nwl_ring_shells_verif}
\end{figure}

Figure~\ref{fig:nwl_ring_shells_verif} shows the local NWL values on each shell as a function of the poloidal angle.
The shape of each shells is parameterized by a center point at a major radius $R_0$ and minor radius as a function of the angle around the center point determined by Fourier modes up to $n=9$.
The residual nonuniformities in the worst case are less than 1.5\%.

\subsection{Surfaces around a tokamak-like plasma geometry}
\begin{figure}[htb]
    \centering
    \includegraphics[trim={0 1.2cm 0 1.5cm},clip,width=0.99\textwidth]{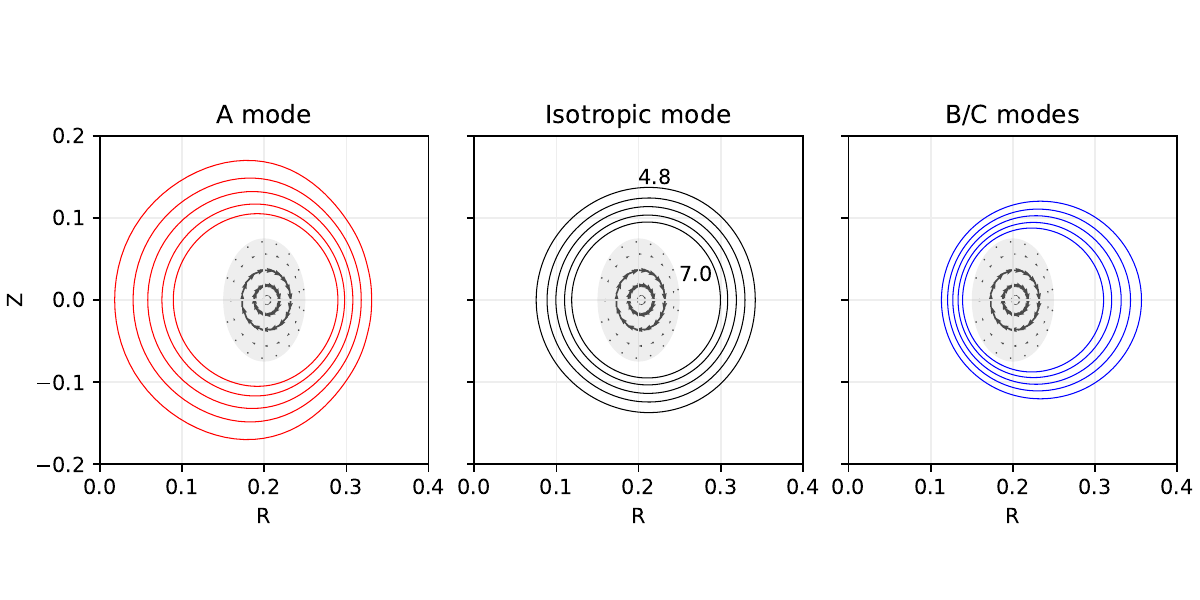}
	\caption{Toroidal shells with a uniform NWL around a tokamak-like source. For each emission mode, the outermost shell has a NWL of 4.8 per unit area and the innermost 7.0 per unit area. The gray ellipse is the cross section of the tokamak-like plasma geometry. Arrows indicate the local direction of the poloidal field and their lengths indicate the strength of emission from each source filament.}
    \label{fig:nwl_tokamaklike_shells}
\end{figure}

As a second demonstration, we construct toroidal shells with uniform NWL load around tokamak-like plasmas with the different spin polarization modes.
The DESC package\cite{paniciDESCStellaratorCode2023} was used to model a tokamak-like plasma geometry with an elliptical outer boundary ($\kappa=1.5$) and a $q$ profile from 1.05 at the axis to $2\pi$ at the edge, and then to compute the $\bc$ field and quadrature weights for a set of 85 points.
This computation uses the formulas for emission from angled (not purely toroidal) fields: $g_{HAa}$, $g_{VAa}$, $g_{HBa}$, and $g_{VBa}$.
The fusion power is assumed to be proportional to $(1 - \hat{\rho}^4)^4$ where $\hat{\rho}$ is the normalized radial coordinate; this creates a flat-topped profile.
As in the previous subsection, the shapes of the shells are chosen to minimize the sum of the squared differences from the target NWL value on each segment.
Target values of 4.8 per unit area to 7.0 per unit area are chosen.
Fig~\ref{fig:nwl_tokamaklike_shells} shows the resulting shells.
Again, the surfaces move inward for the A mode and outward for the B/C modes, though less so than for a single ring source in a toroidal field.
This is because the source has a finite spatial extent, and because the finite $q$-profile means that each toroidal filament has a different angle of peak emission.

%\section{Interior of mirror machine (convex only)}
\section{Additional geometries: the outer exterior of levitated dipole flyer with convex cross section}
\label{sec:dipole}
\begin{figure}[htb]
    \centering
    \includegraphics[width=0.60\textwidth]{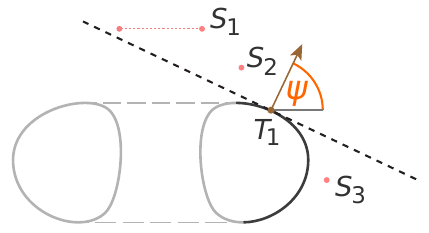}
    \caption{Target point $T_1$ is on the outward-facing part of a levitated dipole flyer magnet. The normal to the surface at its location is $\psic$. The source ring at $S_1$ is entirely visible to $T_1$, $S_2$ is partially visible, and $S_3$ is not visible.}
    \label{fig:dipoleexterior}
\end{figure}

While this paper has been focused on tokamak-like geometries, this penultimate section presents the formula for the limiting integration angles $\pm \phi_m$ for points on the outward-facing portion of a levitating dipole flyer magnet which has a convex cross section.
The outward-facing part of the flyer will receive the highest neutron wall load\cite{simpsonReactorStudyBreakeven2024}.
Figure~\ref{fig:dipoleexterior} shows an example of this geometry, with the outward-facing portion on the right hand side shown with a darker stroke. Point $T_1$ is the example target point and points $S_1 \ldots S_3$ points representing the nearest part of three axisymmetric plasma source rings.
This case is particularly easy, as a target point such as $T_1$ on the outward-facing surface has no sightlines to any other part of the flyer.\footnote{In contrast, the calculation of the limiting angle for \textit{inward}-facing parts of the torus has several special cases, and even for a circular cross section requires solving a third-degree polynomial\cite{skalaLineTorusIntersectionRay2013}.}
Let $\psic$ be the angle between a radial unit vector and the normal at a target point with major radius $r$ on the outside of the flyer ring.
Then $-\cot(\psic)$ is the `inverse slope' (run over rise) of the plane defined by the normal.
Let $p$ be the radius of a ring of plasma with relative height $z$ above the target point.
The limits of integration $\pm\phi_m$ is the angle corresponding to the points on the plasma which intersect the plane defined by the normal, that is:
\begin{equation}
\label{eq:flyer}
\phi_m = \arccos\left(\frac{r - z \cot(\psic)}{p}\right).
\end{equation}
In Figure~\ref{fig:dipoleexterior} the point $S_1$ is a source ring which is entirely visible ($\phi_m = \pi$) to the target point $T_1$.
This is the case if $-p > r - \cot(\psic) z$.
The ring at point $S_2$ is partially visible, since the near and distant points on the ring are on opposite sizes of the plane defining $T_1$'s horizon.
The ring with the near-point $S_3$ is located on the wrong side of $T_1$'s horizon and none of the ring is visible.
This is the case if $p < r - \cot(\psic) z$.
Once the integration angle is determined, one can use Eq.~\ref{eq:emissionsum} (and analogous expressions for polarized sources) to determine the reduced intensity just as in the case for a tokamak-like geometry.

\section{Summary and directions for future work}
This paper develops formulas for the neutron wall load from an axisymmetric spin-polarized fusion ring source, in a restricted set of geometries. 
While the formulas for the most general cases are lengthy, they are comprised of only algebraic expressions, trigonometric functions, and incomplete elliptic integrals, which are well-understood special functions.
Therefore they are fast to evaluate and automatically differentiable, enabling their use in gradient-based optimization schemes.
As an example, employing the \anar{} python library described in this work, we presented a set of toroidal ``first wall'' geometries optimized to have a uniform neutron wall load.

The computation of the intensity consists of two main parts: a set of kernels $g$ which express the reduced intensity from some angular portion of a single (spin-polarized) ring at a patch on the wall, and an algorithm to compute the portion of the ring which is visible from that patch.
The same kernels can be used for \textit{any} axisymmetric geometry, but 
future work is needed to describe the algorithms to compute the visible angle in additional geometries.
Cylindrical, non-convex-cross-section geometries without a center stack, such as that of a mirror machine with end cells, can result in the occlusion of the nearest part of a ring's arc.
Toroidal geometries with non-convex cross sections (such as that of a tokamak with a long-legged divertor) could have occlusions of both near and far parts of a ring's arc.
These geometries could therefore require up to two evaluations of the above formulas to compute the neutron wall load from a plasma filament on a wall patch.
More complicated cross sections which are not single closed curves (such as the entire cross section of a levitated dipole reactor) could require additional evaluations.
Speculatively, it may be possible to evaluate in similar manners more complicated kernels concerning the angular distributions of neutrons impinging on a wall patch (for example, involving higher powers of $\cos\nuc$, or even the azimuthal angle around $\normc$).

%%%%%%%%%%%%%%%%%%%%%%%%%%%%%%%%%%%%%%%%%%
%
% Appendices
%
%%%%%%%%%%%%%%%%%%%%%%%%%%%%%%%%%%%%%%%%%%
\appendix
\section{Definitions of the elliptic integrals}

The formulas presented in this paper are in terms of algebraic functions, trigonometric functions, and elliptic integrals.
These include the incomplete elliptic integrals of the first kind,
$$F(\ellamp, \ellpar) = \int_0^{\ellamp}(1 - \ellpar\sin^2(x))^{-1/2} \, dx,
$$
and incomplete elliptic integrals of the second kind,
$$E(\ellamp, \ellpar) = \int_0^{\ellamp}(1 - \ellpar\sin^2(x))^{1/2} \, dx.
$$
For geometries without occlusion of the ring, complete elliptic integrals of the first kind
$$K(\ellpar) \equiv F(\pi/2, \ellpar)$$
and the complete elliptic integral
$$E(\ellpar) \equiv E(\pi/2, \ellpar)$$
appear in the solution.

\section{Component integrals}
\label{sec:compint}
The integrals in this paper were computed using the \texttt{Rubi} symbolic integration library\cite{richRulebasedIntegrationExtensive2018} within a Mathematica environment.
The symbolic integration algorithms proceed more smoothly when complex integrands are broken up into component parts.
For the integrands under consideration here, most of the complexity is in the sum of terms in a numerator, so it is easy to separate the terms.
All the $E$ and $F$ functions here have \textit{amplitude} $\boldsymbol{\varphi} = \phi_m/2$ and \textit{parameter} $\ellpar = -2/(h-1)$.
The first two integrals, with a power of $3/2$ in the denominator, are used for isotropic emission patterns.
\begin{equation}
\int_{-\phi_m}^{\phi_m}\frac{1}{(h - \cos\phi_m)^{3/2}}\,d\phi = \frac{4 \sin (\phi_m )}{\left(h^2-1\right) \sqrt{h-\cos (\phi_m )}}+\frac{4}{\sqrt{h-1} (h+1)}E + 0\,F
\end{equation}
\begin{equation}
\int_{-\phi_m}^{\phi_m}\frac{\cos\phi}{(h - \cos\phi)^{3/2}}\,d\phi =
\frac{4 h \sin (\phi_m )}{\left(h^2-1\right) \sqrt{h-\cos (\phi_m )}} \\
+\frac{4 h}{\sqrt{h-1} (h+1)}E
-\frac{4}{\sqrt{h-1}}F
\end{equation}
Integrals with a power of $5/2$ in the denominator are used for spin-polarized emission patterns.
\begin{equation}
\begin{split}
\int_{-\phi_m}^{\phi_m}\frac{1}{(h - \cos\phi)^{5/2}}\,d\phi  &=
-\frac{4 \sin (\phi_m ) \left(-5 h^2+4 h \cos (\phi_m )+1\right)}{3\left(h^2-1\right)^2 (h-\cos (\phi_m ))^{3/2}} \\
&+\frac{16 h}{3 (h-1)^{3/2} (h+1)^2}E
-\frac{4}{3 (h-1)^{3/2} (h+1)}F
\end{split}  
\end{equation}

\begin{equation}
\begin{split}
\int_{-\phi_m}^{\phi_m}\frac{\cos\phi}{(h - \cos\phi)^{5/2}}\,d\phi &=
-\frac{4 \sin (\phi_m ) \left(\left(h^2+3\right) \cos (\phi_m )-2\left(h^3+h\right)\right)}{3 \left(h^2-1\right)^2 (h-\cos (\phi_m ))^{3/2}} \\
&+\frac{4 \left(h^2+3\right)}{3 (h-1)^{3/2} (h+1)^2}E
-\frac{4 h}{3 (h-1)^{3/2} (h+1)}F
\end{split}
\end{equation}

\begin{equation}
\begin{split}
\int_{-\phi_m}^{\phi_m}\frac{\cos^2\phi}{(h - \cos\phi)^{5/2}}\,d\phi &=
-\frac{4 h \sin (\phi_m ) \left(h \left(h^2-5\right)-2 \left(h^2-3\right) \cos (\phi_m )\right)}{3 \left(h^2-1\right)^2 (h-\cos (\phi_m ))^{3/2}} \\
&-\frac{8 h \left(h^2-3\right)}{3(h-1)^{3/2} (h+1)^2}E
+\frac{4 \left(2 h^2-3\right)}{3 (h-1)^{3/2} (h+1)}F
\end{split}
\end{equation}

\begin{equation}
\begin{split}
\int_{-\phi_m}^{\phi_m}\frac{\cos^3\phi}{(h - \cos\phi)^{5/2}}\,d\phi &=
\frac{4 h^2 \sin (\phi_m ) \left(\left(5 h^2-9\right) \cos (\phi_m)-4 h \left(h^2-2\right)\right)}{3\left(h^2-1\right)^2 (h-\cos (\phi_m ))^{3/2}} \\
&-\frac{4 \left(8 h^4-15 h^2+3\right)}{3(h-1)^{3/2} (h+1)^2}E
+\frac{4 \left(8 h^2-9\right) h}{3 \sqrt{h-1} \left(h^2-1\right)}F
\end{split}
\end{equation}

\begin{equation}
\begin{split}
\int_{-\phi_m}^{\phi_m}\frac{\sin^2\phi}{(h - \cos\phi)^{5/2}}\,d\phi &= \frac{4 \sin (\phi_m ) \left(h^2-2 h \cos (\phi_m)+1\right)}{3 \left(h^2 - 1\right) (h - \cos (\phi_m ))^{3/2}} \\
&+\frac{8 h}{3 \sqrt{h-1} (h+1)} E
-\frac{8}{3 \sqrt{h-1}} F
\end{split}
\end{equation}

\begin{equation}
\begin{split}
\int_{-\phi_m}^{\phi_m}\frac{\cos\phi \, \sin^2\phi}{(h - \cos\phi)^{5/2}}\,d\phi &= \frac{4 \sin (\phi_m ) \left(4 h^3+\left(3-5 h^2\right) \cos (\phi_m )-2
   h\right)}{3 \left(h^2-1\right) (h-\cos (\phi_m ))^{3/2}} \\
&+\frac{8 \left(4 h^2-3\right)}{3 \sqrt{h-1} (h+1)}E -\frac{32 h}{3 \sqrt{h-1}}F
\end{split}
\end{equation}

\begin{landscape}
\section{Expressions for emission proportional to \texorpdfstring{$\cos^2\theta$}{cosine squared theta} in angled fields}\label{sec:cosdependence}
This appendix lists expressions for the reduced intensity for emission proportional to $\cos^2\thetac$. While these do not correspond to any physical emission pattern from spin-polarized fusion, they are useful for cross-checking with the isotropic, A mode, and B/C mode expressions. 

\subsection{Inboard wall, or horizontal component of a surface's normal}
Dropping terms with odd powers of $\sin\phi$, which by symmetry are zero,
\begin{equation}
\begin{split}
  g_{Rica} = \int_{-\phi_i}^{\phi_i}\,d\phi \frac{\pc \cos\nuc \cos^2\thetac}{|\Deltac|^2} &= \int_{-\phi_i}^{\phi_i} \frac{d\phi}{|\Deltac|^{5}}\Big[
   -p u (p \cos \alpha -z \sin \alpha )^2 \sin ^2\beta * 1 % \\
   +p (p \cos \alpha -z \sin \alpha ) \left(\left(p^2+2 u^2\right) \cos \alpha -p z \sin \alpha \right) \sin ^2\beta * \cos\phi \\
  &-p u \cos \alpha  \left(\left(2 p^2+u^2\right) \cos \alpha -2 p z \sin \alpha \right) \sin ^2\beta * \cos^2\phi \\
  &+p^2 u^2 \cos ^2\alpha \sin ^2\beta *\cos^3\phi
   -p u^3 \cos ^2\beta *\sin^2\phi
   +p^2 u^2 \cos ^2\beta *\cos\phi\sin^2\phi \Big].
\end{split}
\label{eq:gRicasetup}
\end{equation}
This evaluates to
  \begin{equation}
    \begin{split}
  g_{Rica} = & \frac{2\sqrt{p^2-u^2}}{3 p q^2 \sqrt{p^2-u^2+z^2}}\biggl\{ 2 q \left(p^4-p^2 (u^2-2 z^2)+z^2 (u^2+z^2)\right)+\biggl[-3 q z^2 (p^2+u^2+z^2)\\
             & +\left(2 p^8+p^6 (-6 u^2+z^2)-z^2 (u^2+z^2)^3+p^4 (6 u^4-11 u^2 z^2-5 z^4)+p^2 (-2 u^6+11 u^4 z^2+8 u^2 z^4-5 z^6)\right) \cos (2 \alphac )\\
             & -p z \left(5 p^6+(u^2+z^2)^2 (3 u^2+z^2)+p^4 (-7 u^2+11 z^2)-p^2 (u^4+18 u^2 z^2-7 z^4)\right) \sin (2 \alphac )\biggr] \sin ^2(\betac ) \biggr\} \\
             & -\frac{F}{3 p q u \sqrt{(p-u)^2+z^2}}\biggl\{2 q (2 p^2+u^2+2 z^2)+\biggl[-3 q (p^2+u^2+2 z^2)\\
             & +\left((p^2-u^2)^3-2 (p^4-3 p^2 u^2+2 u^4) z^2-5 (p^2+u^2) z^4-2 z^6\right) \cos (2 \alphac )\\
             & +2 p z \left(-2 (p^2-u^2)^2-3 (p^2+u^2) z^2-z^4\right) \sin (2 \alphac )\biggr] \sin ^2(\betac )
    \biggr\} \\
      & +\frac{\sqrt{(p-u)^2+z^2}\,E}{3 p u\,q^2}
      \biggl\{2 q \left(2 p^4+u^4+3 u^2 z^2+2 z^4+p^2 (-3 u^2+4 z^2)\right)+\biggl[-3 q \left((p^2-u^2)^2+3 (p^2+u^2) z^2+2 z^4\right)\\
      &+\left(p^8-p^6 (2 u^2+z^2)-(u^2+z^2)^3 (u^2+2 z^2)+p^2 (u^2+z^2) (2 u^4+15 u^2 z^2-7 z^4)-p^4 (11 u^2 z^2+7 z^4)\right) \cos (2 \alphac )\\
      & +2 p z \left(-2 (p^2-u^2)^2 (p^2+u^2)-5 (p^2-u^2)^2 z^2-4 (p^2+u^2) z^4-z^6\right) \sin (2 \alphac)\biggr] \sin ^2(\betac )\biggr\}
    \end{split}
    \label{eq:gRica}
  \end{equation}
  with $F$ and $E$ as for $\sin^2\thetac$ (see Eq.~\ref{eq:gRiAa}).
Equation~\ref{eq:gRiI} can be expressed as the sum of this expression and Eq.~\ref{eq:gRiAa}.

For an arbitrary integration angle, the leading term of $g_{Rica}$ becomes
\begin{equation}
    \begin{split}
     h_{HcaL}&=\frac{\sin(\phi_m)}{3 q_r^2 \left(p^2+r^2+z^2-2 p r \cos (\phi_m )\right)^{3/2}}\bigg\{
       2 q_r \Bigl(2 p^6+(r^2+z^2)^2 (r^2+2 z^2)+p^4 (r^2+6 z^2)+p^2 (-4 r^4+6 r^2 z^2+6 z^4) \\
     & -p r \Bigl(5 p^4-6 p^2 r^2+r^4+2 (5 p^2+3 r^2) z^2+5 z^4\Bigr) \cos(\phi_m)\Bigr)
      + \biggl[
         3 q_r \Bigl(-p^4 (r^2+2 z^2)-(r^2+z^2)^2 (r^2+2 z^2)\\
     &+2 p^2 (r^4-3 r^2 z^2-2 z^4)
      +p r \left((p^2-r^2)^2+6 (p^2+r^2) z^2+5 z^4\right) \cos (\phi_m )\Bigr)
      +\Bigl(4 p^{10}+4 p^2 z^2 (7 r^2-3 z^2) (r^2+z^2)^2\\
     &-(r^2+z^2)^4 (r^2+2 z^2)+p^8 (-9 r^2+6 z^2)
      +4 p^6 (r^4-7 r^2 z^2-2 z^4)+2 p^4 (r^6-3 r^2 z^4-10 z^6)\\
     & +p r \left(-7 p^8+4 p^6 (5 r^2-2 z^2)+(r^2+z^2)^3 (r^2+5 z^2)+4 p^2 (r^2+z^2) (r^4-15 r^2 z^2+4 z^4)+2 p^4 (-9 r^4+28 r^2 z^2+5 z^4)\right) \cos (\phi_m )\Bigr) \cos(2\alphac) \\
     &-2 p z \Bigl(5 p^8-4 p^6 (r^2-4 z^2)+(r^2+z^2)^3 (5 r^2+z^2)-2 p^4 (r^4+8 r^2 z^2-9 z^4)-4 p^2 (r^6+4 r^4 z^2+r^2 z^4-2 z^6) \\
     &-4 p r \left(2 (p^2-r^2)^2 (p^2+r^2)+5 (p^2-r^2)^2 z^2+4 (p^2+r^2) z^4+z^6\right) \cos (\phi_m )\Bigr)\sin(2\alphac)
      \biggr]\sin^2(\betac)
     \bigg\}.
    \end{split}
    \label{eq:hHcaL}
\end{equation}

\newpage
\subsection{Floor, or vertical component of a surface's normal}
\begin{equation}
\begin{split}
    g_{Rfca} = \int_{-\phi_f}^{\phi_f}\,d\phi \frac{\pc \cos\nuc \cos^2\thetac}{|\Deltac|^2} = \int_{-\phi_f}^{\phi_f} \frac{d\phi}{|\Deltac|^{5}}\Big[&p z (p \cos \alpha -z \sin \alpha )^2 \sin ^2\beta  * 1 +2 p v z \cos \alpha  (z \sin \alpha -p \cos \alpha) \sin ^2\beta  * \cos\phi \\
&+p v^2 z \cos ^2\alpha  \sin ^2\beta  * \cos^2\phi  +p v^2 z \cos ^2\beta * \sin^2\phi \Big].
\end{split}
\label{eq:gRfca}
\end{equation}
This evaluates to
\begin{equation}
    \begin{split}
      & g_{Rfca} = \frac{u \left(\sqrt{p^2-u^2}+\sqrt{-u^2+v^2}\right) z}{3 p q^2 \left(p^2+2 t-2 u^2+v^2+z^2\right)^{3/2}} \biggl\{2 q \Bigl(p^4+\left(v^2+z^2\right)^2+4 \left(t-u^2\right) \left(p^2+v^2+z^2\right)+2 p^2 \left(3 v^2+z^2\right)\Bigr)
       +\biggl[3 q \Bigl(3 p^4+2 p^2 \left(2 t-2 u^2-v^2+z^2\right)\\
       &-\left(v^2+z^2\right) \left(4 t-4 u^2+v^2+z^2\right)\Bigr)+\Big(11 p^8+4 p^6 \left(5 t-5 u^2-4 v^2-2 z^2\right)
       -\left(v^2+z^2\right)^3 \left(4 t-4 u^2+v^2+z^2\right)\\
       &-2 p^4 \left(v^4+14 t z^2+25 z^4-2 u^2 \left(11 v^2+7 z^2\right)+2 v^2 \left(11 t+9 z^2\right)\right)+4 p^2 \left(v^2+z^2\right) \left(2 v^4-13 t z^2-8 z^4+v^2 \left(7 t-6 z^2\right)+u^2 \left(-7 v^2+13 z^2\right)\right)\Big) \cos (2 \alphac )\\
       &+8 p z \left(-4 p^6+2 p^2 (v^2-z^2) \left(3 t-3 u^2+v^2+z^2\right)+\left(v^2+z^2\right)^2 \left(t-u^2+v^2+z^2\right)+p^4 \left(7 u^2+v^2-7 \left(t+z^2\right)\right)\right) \sin (2 \alphac )\biggr] \sin ^2(\betac)
        \biggr\} \\
      &+\frac{z \;F}{3 p q \sqrt{(p-v)^2+z^2}}\biggl\{-2 q+\left(3 q+\left(p^4-2 p^2 \left(v^2-3 z^2\right)+\left(v^2+z^2\right)^2\right) \cos (2 \alphac )+2 p z \left(p^2-v^2-z^2\right) \sin (2 \alphac)\right) \sin ^2(\betac )\biggr\} \\
      &+\frac{z \sqrt{(p-v)^2+z^2}\;E}{3 p\,q^2} \biggl\{2 q \left(p^2+v^2+z^2\right)+\biggl[3 q \left(p^2-v^2-z^2\right)+\left(5 p^6+p^2 \left(7 v^2-13 z^2\right)
   \left(v^2+z^2\right)-\left(v^2+z^2\right)^3-p^4 \left(11 v^2+7 z^2\right)\right) \cos (2 \alphac )\\
      &+2 p z \left(-7 p^4+6 p^2 (v^2-z^2) +\left(v^2+z^2\right)^2\right) \sin (2 \alphac )\biggr] \sin ^2(\betac ) \biggr\}.
    \end{split}
    \label{eq:gRfCa}
\end{equation}

For an arbitrary integration angle, the leading term for $g_{Rfca}$ becomes
\begin{equation}
  \begin{split}
    h_{VcaL} &= \frac{r z \sin (\phi_m)}{3 q_r^2 \left(p^2+r^2+z^2-2 p r \cos (\phi_m)\right)^{3/2}}\biggl\{
2 q_r \Bigl(p^4+(r^2+z^2)^2+2 p^2 (3 r^2+z^2)-4 p r
    (p^2+r^2+z^2) \cos (\phi_m)\Bigr)\\
    & +\biggl[3 q_r \Bigl(3 p^4-2 p^2 ( r^2-z^2)-(r^2+z^2)^2+4 p r
   (-p^2+r^2+z^2) \cos (\phi_m)\Bigr)
     +   \Bigl(11 p^8+8 p^2 (r^2-4 z^2)
   (r^2+z^2)^2-(r^2+z^2)^4\\
    &-8 p^6 (2 r^2+z^2)-2 p^4 (r^4+18 r^2 z^2+25 z^4)+4 p r \left(-5
   p^6+(r^2+z^2)^3+p^4 (11 r^2+7 z^2)+p^2 (-7 r^4+6
   r^2 z^2+13 z^4)\right) \cos (\phi_m)\Bigr)\cos (2 \alphac ) \\
    &-8 p z \left(4 p^6-p^4 (r^2-7 z^2)-(r^2+z^2)^3+2 p^2 (-r^4+z^4)+p r \left(-7 p^4+6 p^2 (r^2-z^2) +(r^2+z^2)^2\right) \cos (\phi_m)\right) \sin (2 \alphac )\biggr] \sin ^2(\betac)
    \biggr\}.
  \end{split}
  \label{eq:hVcaL}
\end{equation}

\end{landscape}

\section*{Acknowledgments}
This work was supported through the U.S. Department of Energy under contract no.\ DE-AC02-09CH11466.

\section*{Data and code availability}
Neutron wall load computations were performed using the \anar{} python library, available at \url{www.github.com/PrincetonUniversity/anarrima}. The scripts used to generate Figures~\ref{fig:nwl_ring_shells}, \ref{fig:nwl_ring_shells_verif}, and \ref{fig:nwl_tokamaklike_shells} are provided with the library as examples.

\bibliography{spfbib}
\bibliographystyle{unsrturl}

\end{document}